%% file: rotation.tex
\PassOptionsToPackage{unicode}{hyperref}
\PassOptionsToPackage{hyphens}{url}
\PassOptionsToPackage{dvipsnames,svgnames,x11names}{xcolor}
\documentclass[
  astrosymb, twocolumn, tighten, twocolappendix]{aastex631}
\usepackage{amsmath,amssymb}
\usepackage{lmodern}
\usepackage{iftex}
\ifPDFTeX
  \usepackage[T1]{fontenc}
  \usepackage[utf8]{inputenc}
  \usepackage{textcomp} 
\else 
  \usepackage{unicode-math}
  \defaultfontfeatures{Scale=MatchLowercase}
  \defaultfontfeatures[\rmfamily]{Ligatures=TeX,Scale=1}
\fi
\IfFileExists{upquote.sty}{\usepackage{upquote}}{}
\IfFileExists{microtype.sty}{
  \usepackage[]{microtype}
  \UseMicrotypeSet[protrusion]{basicmath} 
}{}
\makeatletter
\@ifundefined{KOMAClassName}{
  \IfFileExists{parskip.sty}{%
    \usepackage{parskip}
  }{
    \setlength{\parindent}{0pt}
    \setlength{\parskip}{6pt plus 2pt minus 1pt}}
}{
  \KOMAoptions{parskip=half}}
\makeatother
\usepackage{xcolor}
\usepackage{graphicx}
\makeatletter
\def\maxwidth{\ifdim\Gin@nat@width>\linewidth\linewidth\else\Gin@nat@width\fi}
\def\maxheight{\ifdim\Gin@nat@height>\textheight\textheight\else\Gin@nat@height\fi}
\makeatother
\setkeys{Gin}{width=\maxwidth,height=\maxheight,keepaspectratio}
\makeatletter
\def\fps@figure{htbp}
\makeatother
\input{macros.tex}
\usepackage{mathptmx,txfonts,tikz,bm}
\ifLuaTeX
  \usepackage{selnolig}  
\fi
\usepackage[]{natbib}
\IfFileExists{bookmark.sty}{\usepackage{bookmark}}{\usepackage{hyperref}}
\IfFileExists{xurl.sty}{\usepackage{xurl}}{} 
\hypersetup{
  colorlinks=true,
  linkcolor={Maroon},
  filecolor={Maroon},
  citecolor={xlinkcolor},
  urlcolor={xlinkcolor},
  pdfcreator={LaTeX via pandoc}}

\date{\today}

\begin{document}

\shorttitle{Mixed Rotational Splittings in Red Giants I}
\title{Mode Mixing and Rotational Splittings: I. Near-Degeneracy Effects Revisited}
\input{preamble}
\begin{abstract}
Rotation is typically assumed to induce strictly symmetric rotational splitting into the rotational multiplets of pure p- and g-modes. However, for evolved stars exhibiting mixed modes, avoided crossings between different multiplet components are known to yield asymmetric rotational splitting, particularly for near-degenerate mixed-mode pairs, where notional pure p-modes are fortuitiously in resonance with pure g-modes. These near-degeneracy effects have been described in subgiants, but their consequences for the characterisation of internal rotation in red giants has not previously been investigated in detail, in part owing to theoretical intractability. We employ new developments in the analytic theory of mixed-mode coupling to study these near-resonance phenomena. In the vicinity of the most p-dominated mixed modes, the near-degenerate intrinsic asymmetry from pure rotational splitting increases dramatically over the course of stellar evolution, \edit2{and depends strongly on the mode mixing fraction $\zeta$. We also find that a linear treatment of rotation remains viable for describing the underlying p- and g-modes, even when it does not for the resulting mixed modes undergoing these avoided crossings.} We explore observational consequences for potential measurements of asymmetric mixed-mode splitting, which has been proposed as a magnetic-field diagnostic. Finally, we propose improved measurement techniques for rotational characterisation, \edit2{exploiting the linearity of rotational effects on the underlying p/g modes, while still accounting} for these mixed-mode coupling effects.
\end{abstract}

\hypertarget{introduction}{%
\section{Introduction}\label{introduction}}

Rotation is fundamental to many phenomena in stellar astrophysics. Observationally, rotation alters the surface characteristics of stars, producing both spectroscopic jitter and photometric variability \citep[e.g.][]{santos_surface_2021}. On longer timescales, stellar rotation both generates, and is strongly modified by, the magnetic fields that produce stellar activity cycles, and which vary over their courses \cite[e.g.][]{moss_rotation_1981,loi_topology_2021}. Further into the stellar interior, even slow rotation also induces chemical mixing, and therefore is known to be needed both for deducing the age scale of evolutionary calculations \citep[as in ][]{skumanich_time_1972}, as well as for consistency with the observed chemical abundances of evolved stars \citep[e.g][]{pinsonnealt_evolutionary_1992}. Despite its relative importance in all of these processes, the evolution of rotation off the main sequence remains yet to be fully understood \cite[e.g.][]{aerts_angular_2019}.

As a crucial rung in the methodological ladder undergirding such understanding, asteroseismology permits precise measurements of these rotational dynamics in stellar interiors. 
In nonrotating stars, the oscillation frequencies for nonradial modes of identical degree \(l\) but different azimuthal order \(m\) are degenerate. For sufficiently slow rotation rates, the degeneracy between different \(m\) orders is lifted, producing a multiplet whose mode frequencies are split symmetrically around the zonal mode frequency. \edit1{Ignoring latitudinal dependences,} to first order in perturbation analysis, this splitting is given as
\begin{equation}
    \omega_{nlm} - \omega_{nl,0} \equiv \delta\omega_\text{rot} \sim m \beta_{nl}\int \Omega(r) K_{nl}(r) \mathrm dr,\label{eq:kern}
\end{equation}
where \(K_{nl}\) is an appropriate unimodular rotational kernel. This expression emerges from perturbation analysis of the quadratic Hermitian eigenvalue problem \citep[QHEP:][]{lyndenbell_stability_1967}
\begin{equation}
    \left(\omega^2 \mathcal D + \omega \mathcal R + \mathcal L \right) \bm{\xi} = 0,\label{eq:qhep}
\end{equation}
where \(\mathcal{L}\) is the nonrotating wave evolution operator, \(\bm{\xi}\) is the Lagrangian displacement eigenfunction with eigenvalue \(\omega\), and \(\mathcal{D}\) is the standard inner product. The quantities in \cref{eq:kern} are derived from the diagonal matrix elements (i.e leading-order perturbation) of the Hermitian rotation operator \(\mathcal{R}\), assumed to be heuristically small. Techniques built upon this perturbative construction \citep{hansen_1977, gough_1981,duvall_internal_1984} have been employed to great effect in studying the rotational dynamics of the Sun \citep[][etc.]{howe_rotation_2009,basu_antia_2019} as well as, more generally, main-sequence stars exhibiting solar-like oscillations \citep[][etc.]{schunker_inversion_2016a, schunker_inversion_2016b, Bazot_2019}.

Expressions of \edit1{this} kind have also been applied to evolved stars, and specifically to subgiants and first-ascent red giants \cite[e.g.][]{deheuvels_seismic_2012,mosser_spin_2012, triana_2017, dimauro_2018, gehan_core_2018}. Such stars possess an interior g-mode cavity coupled evanescently to an exterior p-mode cavity \citep{unno_nonradial_1989,ong_surface_1}. Consequently, they exhibit mixed modes, which take on both p-like and g-like character in different parts of the star. Ignoring rotation, the evolution of mixed-mode frequencies is well-approximated by those of pure p- and g-modes, except where they come into resonance, at which points their frequencies exhibit avoided crossings over the course of stellar evolution \citep[e.g.][]{deheuvels_insights_2010, bedding_replicated_2012} due to the coupling between the two cavities. These avoided crossings notwithstanding, the rotational splittings of these mixed modes are nonetheless often assumed to be well-described by expressions of the form of \cref{eq:kern}. Over the course of their post-main-sequence evolution, the radiative cores of these stars shrink, and their envelopes expand simultaneously, resulting in significant radial differential rotation were angular momentum to be conserved. Phenomenologically, this has been approximated using a two-zone model of radial differential rotation \citep[e.g.][]{klion_diagnostic_2017}, where the core and envelope are assumed to rotate as solid bodies, but at separate rates. Applied to this scenario, \cref{eq:kern} suggests that the rotational splitting width of these mixed modes may be approximated as linear combinations of some notional core and envelope rotational rates:
\begin{equation}
    \delta\omega_{\text{mixed}, i} \sim m \left(\zeta_i \beta_\text{core}\Omega_\text{core} + (1 - \zeta_i) \beta_\text{env}\Omega_\text{env}\right),\label{eq:zetarot}
\end{equation}
where \(\zeta_i\) is a mixing fraction associated with the \(i^\text{th}\) mixed mode, which is close to 1 for a g-dominated mode and close to 0 for a p-like one.

In general, however, first-order perturbation theory may not correctly describe mixed-mode rotational splittings. In particular, \citet[][hereafter \dheuv{}]{deheuvels_near_2017} demonstrate asymmetric rotational splitting in subgiants (\(\Delta\nu \sim 30\ \mu\)Hz) to emerge from both analytic considerations and explicit frequency calculations for a series of stellar models. This phenomenon arises when the rotational splitting is comparable in magnitude to the mode coupling strength, in which case the differential mode bumping experienced by the different azimuthal components of near-degenerate mixed-mode multiplets, where each multiplet component executes an avoided crossing at a slightly different age, results in an effective asymmetric splitting. \dheuv{} also note that in these subgiants, the coupling strengths for dipole modes are so strong that the asymmetry in the dipole multiplet splitting ceases to be a concern; they consequently limit their attention to quadrupole modes. However, the mixed-mode coupling strengths are known to decrease rapidly as a star ascends the red giant branch. Dipole modes in more evolved red giants are consequently much more susceptible to level-crossing-induced multiplet asymmetry for on-resonance p/g-mode pairs. At the same time, the approximate nonperturbative construction \edit1{of} \dheuv, which considers only the case of two interacting modes, \edit1{was developed specifically to suit the restricted scope of} subgiants and young red giants, where individual avoided crossings can be identified and treated in isolation. The relative density of g-modes to p-modes increases dramatically with evolution, however, in which case analysis of this phenomenon ceases to be tractable by reduction to the two-mode system. \edit1{Closed-form analytic solutions of the kind considered in \dheuv{} become impossible in this regime, and} so far it has only been examined by way of brute-force numerical solutions \citep[e.g.][]{ouazzani_rotation_2013}. \edit1{For these reasons, \dheuv{} defer a detailed analysis of these near-degeneracy effects in more evolved red giants to a later work. Our aim in this work is to fill this lacuna.}

We will show that in general, the validity of first-order expressions like \cref{eq:zetarot} strongly depends on the choice of basis functions used to describe the Lagrangian fluid displacements, \(\bm{\xi}\), of these oscillations. While they admit a natural description as the orthogonal eigenfunctions generated by \(\mathcal{L}\) and \(\mathcal{D}\), mixed modes are also well-described as linear combinations of purely p-like ``\(\pi\)-modes'' and purely g-like ``\(\gamma\)-modes'', in the sense of \citet[][hereafter \ob]{aizenman_avoided_1977,ong_semianalytic_2020}. \edit1{These numerical $\pi$ and $\gamma$-modes serve to approximate the notional pure p- and g-modes underlying these avoided crossings}. Heuristically, asymmetry in the rotational splitting is the ``local'' difference in the frequency perturbation owing to mode coupling, across a frequency range set by the size of the rotational splitting. Accordingly, this asymmetry is a second-order effect, which cannot be described by first-order perturbation analysis. So too, however, is the coupling between the two mode cavities that gives rise to these avoided crossings in the first place. We show in particular that the rotationally-split \(\pi\) and \(\gamma\)-mode multiplets are separately well-described by the symmetric splitting predicted by first-order perturbation analysis, even if the coupled mixed mode multiplets may not be.

These asymmetric splittings also present significant methodological difficulties, in addition to these conceptual ones. Both the operators \(\mathcal{L}\) and \(\mathcal{D}\) of \cref{eq:qhep} are diagonal by definition in the natural basis of mixed-mode eigenfunctions. However, \dheuv{} show that these avoided crossings, which yield asymmetric rotational splittings in the first place, mathematically require nonzero matrix elements off the diagonal in this eigenvalue problem; these must necessarily be attributed to the rotation operator. Asymmetric splittings are thus ipso facto incompatible with the purely diagonal rotation matrices which are ordinarily assumed. As we will see, existing asymptotic approximations to asymmetric splitting \citep[e.g.][]{goupil_seismic_2013, mosser_period_2015} do not return correct expressions for these off-diagonal elements. Ultimately, more sophisticated techniques for dealing with stellar rotation \citep[e.g.~rotational inversions, as attempted for mixed modes in][]{ahlborn_rotation_2020, ahlborn_improved_2022, fellay_asteroseismology_2021} also rely on this diagonal property of these rotational matrices. The validity of inversion techniques when this assumption fails has hitherto not been well examined, and in fact has come into question of late \citep[e.g.][]{bellinger_inference_2021}. Conversely, the recovery of a set of basis functions for which the approximation of diagonality holds well will be necessary if any attempt at rotational inversion using mixed modes were to be well-posed.

In this work, we make use of the new analytic developments described in \ob{} to examine these near-degeneracy effects in the context of evolved red giants, where the individual avoided crossings are not easily described by reduction to the two-state system. In particular, the description of \dheuv{} was limited to the natural basis of mixed-mode eigenfunctions. Insight into how rotational splitting related to mode coupling was obtained implicitly, by truncating the domains of integral expressions of the same kind as \cref{eq:kern} ad hoc. Such an approach implicitly assumes \edit1{that exactly one notional p-mode couples to exactly one notional g-mode}, and does not easily generalise to more general configurations, such as the many-to-one mode coupling seen in the mixed dipole modes of evolved red giants. We \edit1{approximate these notional pure modes by performing} calculations in the basis of isolated \(\pi\) and \(\gamma\) modes, supplying the elements of the coupling matrices in \cref{eq:qhep} with direct reference to stellar structure by way of the integral expressions in \ob{} and \citet{ong_surface_1}. In this manner, we render the many-mode problem tractable, both analytically and numerically, without resorting to brute-force techniques; this also accounts for nonlocal coupling, unlike other constructions using a mixing function \(\zeta\). In \autoref{sec:construction}, we describe the linear-algebraic construction used for subsequent calculations, present some analytic \edit1{limits} on the asymmetry in the rotational splitting from various considerations, and compare these to existing expressions in the literature. In \autoref{sec:numerics} we present numerical results from applying this construction to evolutionary models of red giants, and we consider extensions to currently-used techniques in \autoref{sec:prospects}. We summarise our findings in \autoref{sec:conclusion}.

\hypertarget{algebraic-construction}{%
\section{Algebraic Construction}\label{algebraic-construction}}

\label{sec:construction}

We formalise our intuition that the rotation operator appearing in \cref{eq:qhep} is weak, by formulating this QHEP as a perturbation to the standard Hermitian eigenvalue problem:
\begin{equation}
    \left(\omega^2 \mathcal{D} + \lambda \omega \mathcal{R} + \mathcal{L} + \lambda^2\mathcal{V} + \mathcal{O}(\lambda^3) \right)\bm{\xi} = 0, \label{eq:qhep:lambda}
\end{equation}
where \(\lambda \in [0,1]\), which reduces to the usual Hermitian eigenvalue problem in the nonrotating case as \(\lambda\to0\). The power of \(\lambda\) assigned to each operator is chosen to coincide with the highest power of \(m\) on which it depends, which we will find useful later. While typically only the first-order term \(\mathcal{R}\) is considered in analyses of rotational splitting, a complete accounting of rotational effects should in principle also include the further perturbation \(\mathcal{V}\) to the wave operator, which accounts for both dynamical and structural effects. The former results from centrifugal forces in the corotating frame, while the latter is a consequence of rotation also deforming the stellar structure with respect to its nonrotating configuration. Since both phenomena are axially symmetric, their effects on the mode frequencies are even in \(m\) (thereby providing a further asymmetric component in the rotational splitting). However, these only enter into the wave equation to second order in \(\Omega\) \citep[cf.][]{aertsbook}. We will consider only second-order dynamical effects for the purposes of comparison. As we will show, the omission of higher-order structural effects does not materially change our conclusions.

We relate these operators to matrices by taking inner products with respect to some set of basis wavefunctions. In general, these wavefunctions have radial, poloidal, and toroidal components, and, after separation of variables, may be written in slowly rotating stars as linear combinations of vector spherical harmonics \citep{arfken_weber_2011} as
\begin{equation}
    \bm{\xi} = \xi_r \mathbf{Y}_l^m + \xi_t\mathbf{\Psi}_l^m + \xi_h \mathbf{\Phi}_l^m.
\end{equation}
We assume these basis functions to be derived from standard methods (e.g.~they may be the eigenfunctions obtained by solving \cref{eq:qhep:lambda} with \(\lambda\to 0\)). For a differential operator \(\mathcal{Q}\), there is a corresponding matrix \(\mathbf{Q}\) associated with this set of basis functions. We may find the elements of this matrix by taking integrals over the equilibrium structure as
\begin{equation}
    Q_{ij} = \left<\bm{\xi}_i, \mathcal{Q}\bm{\xi}_j\right> = \int \mathrm d m\ \bm{\xi}_i^* \cdot \mathcal Q [\bm{\xi}_j],
\end{equation}
where the basis functions themselves are assumed to be normalised as \(D_{ii} = \left<\bm{\xi}_i, \bm{\xi}_i\right> = 1\); we make no assumption of orthogonality, instead leaving this information to be described by the matrix \(\mathbf{D}\). Where needed, we will distinguish between these matrices as evaluated against different sets of basis functions using superscripts denoting which basis set has been used. When enough of these matrix elements have been evaluated, the eigenvalues \(\omega\) of \cref{eq:qhep:lambda} may be well approximated by solving the corresponding matrix equation:
\begin{equation}
    \left(\omega^2 \mathbf{D} + \lambda \omega \mathbf{R} + \mathbf{L} + \lambda^2\mathbf{V} + \mathcal{O}(\lambda^3) \right)\mathbf{c} = 0, \label{eq:qhep:matrix}
\end{equation}
where the eigenvectors \(\mathbf{c}\) specify linear combinations of these basis vectors. Explicit expressions for \(\mathbf{L}\) and \(\mathbf{D}\) were developed in in \ob{} (e.g. their eqs. 31, 39, 40) and \citet{ong_surface_1}. We now focus on the perturbing matrices \(\mathbf{R}\) and \(\mathbf{V}\).

\hypertarget{rotation-matrices-and-integral-kernels}{%
\subsection{Rotation matrices and integral kernels}\label{rotation-matrices-and-integral-kernels}}

The primary contribution to the rotational splitting comes from the operator \(\mathcal R\), obtained by a combination of the term in the momentum equation corresponding to the Coriolis force,
\begin{equation}
    \mathbf{f}_\text{cor} = -2 \rho \bm{\Omega} \times \mathbf{v},
\end{equation}
and a change of coordinates from the corotating to the inertial reference frame. The matrix elements of this operator are given by a generalisation of the usual expression \citep[e.g.][]{aertsbook}:
\begin{equation}
\begin{aligned}
    \left<\bm{\xi}_i , \mathcal R\ \bm{\xi}_j \right> &\equiv R_{ij} \\&= 2 m \int \mathrm d r \ \Omega(r)\  r^2 \rho_0 \left(\xi_{r, i} \xi_{r, j} + [l(l+1) - 1] \xi_{t, i} \xi_{t, j} - \xi_{r, i} \xi_{t, j} - \xi_{t, i} \xi_{r, j}\right)\\
    &\equiv 2 m \beta_{ij} \int \mathrm d r\ \Omega(r) K_{ij}(r),\label{eq:rotkernel}
\end{aligned}
\end{equation}
where the kernel \(K_{ij}\) is defined to satisfy \(\int \mathrm d r \ K_{ij}(r) = 1\), such that for uniform rotation we have \(R_{ij} = 2 m \beta_{ij} \Omega\). It is trivial to verify that the diagonal entries of the matrix \(R\) are twice the right-hand-side of \cref{eq:kern}.

We also compute the dynamical frequency shifts induced by the centrifugal force in the corotating reference frame, which is known to scale as \(\Omega^2\) \citep[cf.][]{gough_rotation_1990, kjeldsen_rotation_1998}. Following \citet{lyndenbell_stability_1967}, the wave operator \(\mathcal L\) is subjected to a small perturbation as \(\mathcal L + \lambda^2 \mathcal V\), where
\begin{equation}
\begin{aligned}
    \bm{\xi}_i \cdot \mathcal V \bm{\xi}_j  &= -\bm{\xi}_i \cdot \bm{\Omega} \times (\bm{\Omega} \times \bm{\xi}_j) \\&= (\bm{\Omega} \times \bm{\xi}_i) \cdot (\bm{\Omega} \times \bm{\xi}_j) = \Omega^2 \bm{\xi}_i \cdot \bm{\xi}_j - (\bm{\Omega} \cdot \bm{\xi}_i)(\bm{\Omega} \cdot \bm{\xi}_j).
\end{aligned}
\end{equation}
Integrating over spherical harmonics yields a kernel integral of the form
\begin{equation}
\begin{aligned}
\left<\bm{\xi}_i , \mathcal V \bm{\xi}_j\right> &\equiv V_{ij} = \int \mathrm d r \ \Omega(r)^2  \rho_0 r^2 \left(\xi_{r,i} \xi_{r,j} \cdot {2 \over 3}\left[1 - Q_{lm2}\right] \right. \\
& \left.+ \xi_{h,i} \xi_{h,j} \left[L^2 - {2 \over 3}(L^2 - 3)(1 - Q_{lm2}) + m^2\right] - Q_{lm2}(\xi_{r,i} \xi_{h,j} + \xi_{h,i} \xi_{r,j})\right)\\
&\equiv \gamma_{m, ij} \int \mathrm d r\ \Omega(r)^2 J_{m, ij}(r),\label{eq:centkernel}
\end{aligned}
\end{equation}
where the constants \(\gamma_{m, ij}\) and kernels \(J_{m, ij}\) are defined analogously to \(\beta\) and \(K\) for the first-order rotational splittings: for uniform rotation we have \(V_{ij} = \gamma_{m, ij} \Omega^2\). Here we have used a compact notation for the integral of three spherical harmonics \citep[following][]{gough_rotation_1990} as
\begin{equation}
    Q_{lmn}= \left.\int_{-1}^1 P_n(x)\left[P_{l}^m(x)\right]^2 \mathrm d x \over \int_{-1}^1 \left[P_{l}^m(x)\right]^2 \mathrm d x \right.;\ \ \ Q_{lm2} = {l(l+1) - 3m^2 \over (2l-1)(2l+3)}.
\end{equation}
The omission of structural effects enters into this term. Accounting for them fully would change the precise values of \(\gamma_{ij,m}\) and the structure of the integral kernels \(J\), but does not otherwise substantively modify our subsequent discussion.

\subsection{Perturbation Analysis for Mixed Modes}
\label{subsec:perturbation}

With these definitions in hand, we are now in a position to examine the properties of the rotational splitting in more detail. Supposing that we have numerical access to these matrix elements in the natural basis of mixed-mode eigenfunctions, we may extend the first-order expression, \cref{eq:kern}, to include higher-order terms in perturbation theory. We adopt the usual procedure \citep[e.g.][]{landau_quantum_1965} of expanding the perturbed eigenvalues and eigenfunctions of \cref{eq:qhep:lambda} as an asymptotic series in powers of \(\lambda\):
\begin{equation}
    \begin{aligned}
    \omega_i(\lambda) &= \omega_{i,0} + \lambda \omega_{i,1} + \lambda^2 \omega_{i,2} + \ldots \\
    \bm{\xi}_i(\lambda) &= \bm{\xi}_{i,0} + \lambda \bm{\xi}_{i,1} + \lambda^2 \bm{\xi}_{i,2} + \ldots \\
    \end{aligned}\label{eq:expand}
\end{equation}
Inserting these expressions into \cref{eq:qhep:lambda} and grouping terms by powers of \(\lambda\) yields at last that
\begin{equation}
\begin{aligned}
    \omega_i(\lambda) &= \omega_{i,0} + {\lambda\over 2}R_{ii} \\&+ {\lambda^2\over2\omega_{i,0}}\left(- V_{ii} + {1\over4}R_{ii}^2+\omega_{i,0}^2 \sum_{j \ne i} {\left|R_{ij}\right|^2\over\omega_{i,0}^2 - \omega_{j,0}^2}\right) + \mathcal O(\lambda^3),\label{eq:perturb}
\end{aligned}
\end{equation}
while the eigenfunctions themselves are also perturbed as
\begin{equation}
    \bm{\xi}_i(\lambda) = \bm{\xi}_{i,0} - \lambda \omega_{i,0} \sum_{j \ne i}{R_{ji} \over \omega_{i,0}^2 - \omega_{j,0}^2} \bm{\xi}_{j,0} + \mathcal{O}(\lambda^2).
\end{equation}

Since the rotation matrix elements, specified by \cref{eq:rotkernel}, are all odd in \(m\), and those for second-order effects, \cref{eq:centkernel}, are even in \(m\), the frequencies of the prograde and retrograde components of the multiplet, \(\omega_{m = \pm l}(\lambda)\), may be approximately computed by changing the sign of \(\lambda\), rather than of \(m\): \(\omega_\pm(\lambda) \sim \omega_+(\pm\lambda)\). We can use this property to greatly simplify the dimensionless asymmetry parameter defined in \dheuv{}:
\begin{equation}
    \psi = {\omega_+ + \omega_- - 2 \omega_0 \over \omega_+ - \omega_-}. \label{eq:asym}
\end{equation}
Comparing \cref{eq:asym} with \cref{eq:expand,eq:perturb} gives us the approximate expression
\begin{equation}
    \psi_i \sim \lambda{\omega_{i,2} \over \omega_{i,1}} \sim \left.{\lambda \over 2}\left( \left. {\partial^2 \omega_i \over \partial \lambda^2} \right/{\partial \omega_i \over \partial \lambda} \right) \right|_{\lambda\to0} + \mathcal{O}(\lambda^3),\label{eq:approxasym}
\end{equation}
which we will \edit2{use} extensively in our following calculations. \cref{eq:approxasym} makes explicit the fact that asymmetric splitting arises from second- and higher-order effects. Finally, we will also characterise the systematic error between the true width of the rotational splitting, and that predicted from the first-order expression, \(\omega_{i,1}\). We define a relative error parameter
\begin{equation}
    {\delta \omega_\text{rot} \over \omega_\text{rot}} \equiv \epsilon_i = {\omega_{i,+} - \omega_{i, -} \over 2 \omega_{i, 1}} - 1. \label{eq:dev}
\end{equation}
Inserting \cref{eq:expand} into \cref{eq:dev} yields
\begin{equation}
    \epsilon_i \sim \lambda^2 {\omega_{i,3} \over \omega_{i, 1}} \sim \left.{\lambda^2 \over 3!}\left( \left. {\partial^3 \omega_i \over \partial \lambda^3} \right/{\partial \omega_i \over \partial \lambda} \right) \right|_{\lambda\to0} + \mathcal{O}(\lambda^4);\label{eq:approxdev}
\end{equation}
that is to say, changes to the widths of the rotational multiplets (ignoring asymmetry) only occur to third and higher order in perturbation theory.

It is readily apparent that \cref{eq:kern} follows from truncating \cref{eq:perturb} to first order in \(\lambda\). In general, the higher-order terms in this expansion are potentially dominated by two different kinds of expressions. If the matrices \(R\) and \(V\) are diagonally dominated, then the rotational splitting may be approximated solely in terms of these diagonal elements. Since the asymptotic expansion \cref{eq:perturb} is usually assumed to converge, the multiplet widths are likewise usually assumed to be very well approximated by the diagonal elements of the rotation matrices. Conversely, if the off-diagonal elements cannot be neglected, then pairs of modes close to resonance will have their higher-order terms dominated by powers of sums over resonance factors, \(\sum_{i\ne j} {\left|R_{ij}\right|^2 \left/\left(\omega_{i,0}^2 - \omega_{j,0}^2\right)\right.}\). In the presence of near-degenerate resonant mode pairs, this asymptotic expansion in fact blows up, and we should likewise expect these higher-order effects to become significant near an avoided crossing. Both of these limiting cases yield \edit1{limiting values of $\psi$ and $\epsilon$} for different sources of asymmetry in the rotational splitting, which we will examine separately in the next few subsections.

\edit1{Finally, while accidental degeneracy between modes of the same $l$ and $m$ is not possible in the case of mixed modes (so that these resonance factors never truly become singular), existing analytic treatments of accidental deneracy may still be approximately applied to sets of mixed modes which are merely near degeneracy. We provide a sketch of this procedure for illustrative purposes. Suppose that we may identify a subspace $D_k$ spanned by modes which are near-degenerate \citep[e.g. by the near-degeneracy criterion of][]{lavely_effect_1992} at nonrotating frequency $\omega_{k,0}$. Rather than using the natural mixed-mode basis functions within this subspace, the perturbative expansion is performed instead with respect to linear combinations of mixed modes, $\bm{\eta}_i = \sum_{j} c_{ij} \bm{\xi}_j$, that diagonalise the perturbation (in this case rotation) operator as restricted to within this subspace. Thus, $D_k = \mathrm{span}\left(\left\{\bm{\eta}_i\right\}\right)$, such that $\left<\bm{\eta}_i, \mathcal{R}\bm{\eta}_j\right> \propto \delta_{ij}$. The perturbative series is then developed as above, with the sum over resonance factors restricted to modes outside this subspace. For instance, we will have
\begin{equation}
    \bm{\eta}_i(\lambda) = \bm{\eta}_{i,0} - \lambda \omega_{k,0} \sum_{j\text{ s.t. }\bm{\xi}_j \notin D_k}{R_{ji} \over \omega_{k,0}^2 - \omega_{j,0}^2} \bm{\xi}_{j,0} + \mathcal{O}(\lambda^2)   
\end{equation}
for the perturbed eigenfunctions. We note that if the modes spanning $D_k$ are not actually accidentally degenerate, the wave operator $\mathcal{L}$ is not diagonal with respect to this basis. We will return to these properties in our later discussion.}

\hypertarget{bare-modes}{%
\subsection{Bare modes}\label{bare-modes}}

\edit1{We now consider limiting values of the asymmetry parameter $\psi$ under various configurations of rotation splitting}. Let us first consider the case where the eigenvalues from \cref{eq:qhep} are evaluated when the off-diagonal entries are ignored (i.e.,~bare modes with no rotational coupling), which is usually assumed to be permissible in most treatments of rotational splitting in the literature --- this amounts to assuming that the matrix \(R\) is diagonally dominant. For each mode \(i\) with nonrotating frequency \(\omega_{i,0}\), the rotationally-split mode frequencies can be found as the solutions to the algebraic equation
\begin{equation}
    \omega^2 + 2 \omega  \Omega m b_{ii} - \omega_{i,0}^2 + \Omega^2 g_{m, ii} = 0, \label{eq:diag}
\end{equation}
where we define \(b_{ij}\) and \(g_{ij}\) to be dimensionless quantities such that \(R_{ij} = 2 m \Omega b_{ij}\) and \(V_{ij} = \Omega^2 g_{ij}\) for some effective rotation rate \(\Omega\); for solid-body rotation, \(b_{ij} \to \beta_{ij}\) and \(g_{ij} \to \gamma_{ij}\). The solutions to \cref{eq:diag} are
\begin{equation}
\begin{aligned}
    |\omega_{\pm}| &= \sqrt{\omega_{i,0}^2 + \Omega^2 \left(-g_{m, ii} + m^2 b_{ii}^2\right)} \pm m \Omega b_{ii} \\& \equiv \sqrt{\omega_{i,0}^2 + 2\omega_{i,0}\omega_{i,2}} \pm \omega_{i, 1}. \label{eq:convergent}
\end{aligned}
\end{equation}
On the other hand, the \(m=0\) mode is also perturbed by the centrifugal force, yielding
\begin{equation}
    \omega_0 = \sqrt{\omega_{i,0}^2 - \Omega^2 g_{0, ii}}.
\end{equation}
The asymmetry parameter, \cref{eq:asym}, can be expanded in powers of \(\Omega / \omega_0\) to give
\begin{equation}
\begin{aligned}
    \psi_{i, \text{dyn}} &\sim { \sqrt{\omega_0^2 + \Omega^2 \left(-g_{m, ii} + g_{0, ii} + m^2 b_{ii}^2\right){}} - \omega_0 \over m \Omega b_{ii}} 
    \\&\sim {\Omega \left(-g_{m, ii} + g_{0, ii} + m^2 b_{ii}^2\right) \over 2mb_{ii} \omega_0} + \mathcal{O}\left({(\Omega/\omega_0)}^3\right). \label{eq:bareasym}
\end{aligned}
\end{equation}
This expression, which we can verify satisfies \cref{eq:bareasym}, serves as \edit1{an estimate of} the amount of asymmetry we expect to obtain, resulting purely from second-order dynamical effects. It is small if the condition for slow rotation (\(\Omega \ll \omega\)) is satisfied. We note, furthermore, that \cref{eq:convergent} is even in \(\Omega\) above first order, and thus \(\epsilon\), \cref{eq:approxdev}, vanishes to all orders in \(\lambda\): when the off-diagonal elements of \(R\) are neglected, the width of the splitting is equal to that given by the first-order expression.

\hypertarget{two-state-avoided-crossing}{%
\subsection{Two-state avoided crossing}\label{two-state-avoided-crossing}}

In the opposite extreme, we may also estimate the amount of multiplet asymmetry when the rotational splitting is dominated entirely by near-degeneracy effects (i.e.~the resonance factors in \cref{eq:perturb}). We consider a single \(\pi\)-mode multiplet undergoing an avoided crossing with a single \(\gamma\)-mode multiplet --- this is the scenario also considered in \dheuv{} under the assumption of two-zone differential rotation, although we treat it with a substantially different set of analytic tools. In our subsequent discussion, we will also use this two-zone model of differential rotation, with an angular frequency of \(\Omega_\text{env}\) in the envelope, and \(\Omega_\text{core} = C \Omega_\text{env}\) in the core. The use of the isolated \(\pi\) and \(\gamma\) modes has the advantage that, since the isolated eigenfunctions are already limited in physical extent, the use of \cref{eq:rotkernel} without modifications, with respect to a piecewise-constant function describing \(\Omega(r)\), already very well approximates this two-zone model without requiring ad hoc truncation of the integral limits as in \dheuv.

We generalise the construction of \ob{} to include rotation matrices, yielding a quadratic eigenvalue problem written out with respect to the coefficients of the \(\pi\) and \(\gamma\) basis functions:
\begin{equation}
    \left(\omega^2
    \begin{bmatrix}
    1 & D \\ D & 1
    \end{bmatrix}
    + 2 \omega m \Omega_\text{env}
    \begin{bmatrix}
    \beta_{\pi\pi} & 0 \\ 0 & C \beta_{\gamma\gamma}
    \end{bmatrix}
     -
     \begin{bmatrix}
     \omega_\pi^2 & \alpha \\ \alpha & \omega_\gamma^2
     \end{bmatrix}
     +\Omega_\text{env}^2
     \begin{bmatrix}
     \gamma_{m,\pi} & 0 \\ 0 & C^2 \gamma_{m,\gamma}
     \end{bmatrix}\right) \mathbf{c}
      = 0,
\end{equation}
where the structural coupling factors \(\alpha\) and \(D\), which are the off-diagonal elements of \(\mathcal{L}\) and \(\mathcal{D}\), are specified by overlap integrals between the \(\pi\)- and \(\gamma\)-mode eigenfunctions:
\begin{equation}
\alpha = \int \mathrm{d}m\  \bm{\xi}_\gamma^* \cdot (N^2  \mathbf{e}_r \otimes \mathbf{e}_r + \omega_\pi^2) \cdot \bm{\xi}_\pi,\text{ and } D = \int \mathrm{d}m\  \bm{\xi}_\gamma^* \cdot \bm{\xi}_\pi.\label{eq:overlap}
\end{equation}
\edit1{Again, we have assumed that the off-diagonal elements of the rotational matrix may be ignored. Like $\alpha$ and $D$, they are expressed as overlap integrals between the $\pi$ and $\gamma$ modes, which are small if the two mode cavities are well separated. However, while $\alpha$ and $D$ enter into the problem independently of the rotation rate, the off-diagonal elements of the rotation matrix are not only small from these structural concerns, but are also multiplied by the additionally small rotation rate, $\Omega_e$ --- and then only their squares enter into the expressions for the mode frequencies at second order, per \cref{eq:perturb}. Accordingly, their effects are of a much higher order of smallness than the coupling between the mode cavities, or even second-order rotational effects, and may thus be safely neglected for this analysis.}

\begin{figure}
\centering
\includegraphics{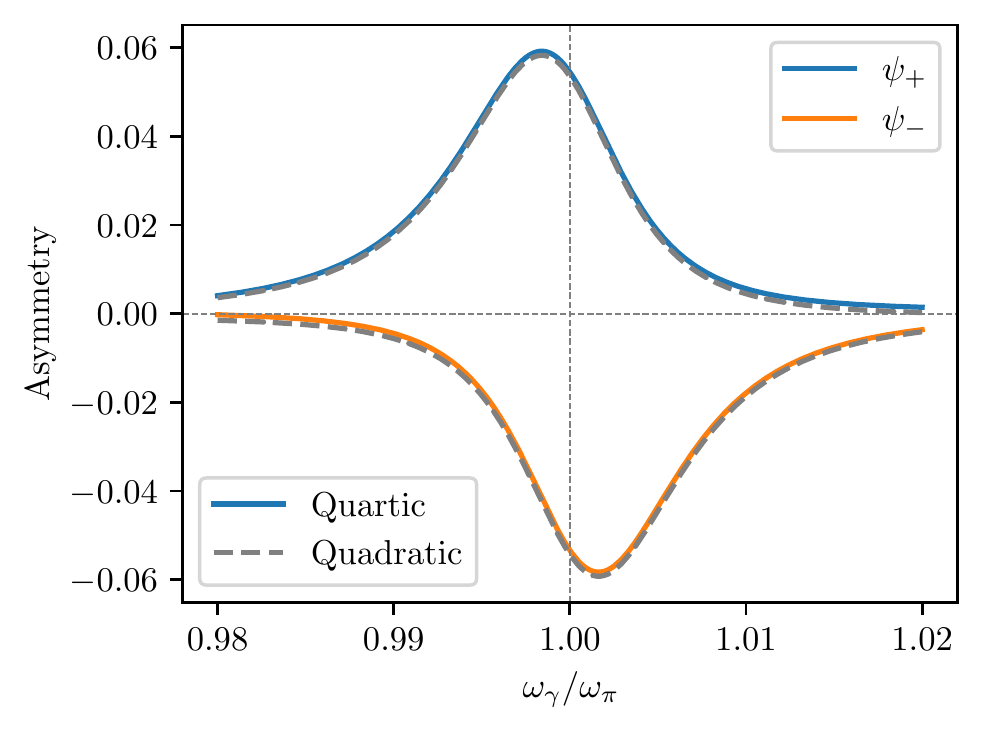}
\caption{Asymmetry parameters \(\psi_\pm\) as a function of \(\omega_\gamma/\omega_\pi\) for a dipole mixed-mode pair, using parameters supplied from a MESA evolutionary model (Model 1, described later). Values derived using the quadratic approximation of \citet{deheuvels_near_2017} are shown with the dashed lines, and are in very good agreement with the solutions to the quartic equations.\label{fig:asym-curves}}
\end{figure}

The mixed modes themselves are linear combinations of the \(\pi\) and \(\gamma\) modes, whose coefficients are specified by the eigenvectors \(\mathbf{c}\).
\edit1{Solutions to this eigenvalue problem can be found in closed form from the roots of the characteristic equation, which is in principle a fourth-order algebraic equation in \(\omega\)}. In practice, however, these expressions are extremely unwieldy (consisting of many layers of nested radicals). We instead solve the approximate problem (as in \dheuv{})
\begin{equation}
    \left(\omega^2
    \begin{bmatrix}
    1 & D \\ D & 1
    \end{bmatrix}
    + 2 m \Omega_\text{env}
    \begin{bmatrix}
    \omega_\pi \beta_{\pi\pi} & 0 \\ 0 & C \beta_{\gamma\gamma} \omega_\gamma
    \end{bmatrix}
     -
     \begin{bmatrix}
     \omega_\pi^2 & \alpha \\ \alpha & \omega_\gamma^2
     \end{bmatrix}
     \right) \mathbf{c}
      = 0,
\end{equation}
with solutions found instead from a quadratic equation in \(\omega^2\), and where we have ignored second-order dynamical effects. Denoting the roots of \edit1{the characteristic} equation by \(\omega^2_\pm\), the asymmetry parameters \(\psi_\pm\) may then be either calculated fully using \cref{eq:asym}, or approximated using \cref{eq:approxasym}. We compare the closed-form quadratic and quartic solutions for \(\psi\), both evaluated via \cref{eq:asym}, in \cref{fig:asym-curves}, for values of these parameters supplied from a red giant evolutionary model (which we describe in more detail in the next section). Since we find them to be in excellent agreement, we will restrict our attention to expressions derived from the much simplified quadratic approximation.

We now make use of certain known properties of \(\pi\) and \(\gamma\) modes. Since they behave like pure p- and g-modes, we may demand \(\beta_\pi \sim 1,\; \beta_\gamma \sim 1 - {1 / l(l+1)}\) at low degree. From \cref{fig:asym-curves} we also see that near-degeneracy effects yield large asymmetries near resonance; accordingly we consider modes at resonance, so that \(\omega_\pi = \omega_\gamma = \omega\). Finally, since we are considering red giants in particular, we take the limit of weak coupling between the p- and g-mode cavities: \(D < {\alpha / \omega^2} \ll 1\). \cref{eq:approxasym} then gives
\begin{equation}
    \psi \sim {1 \over 4}\left(m \Omega_\text{env} \omega \over \omega^2 \mp \alpha\right)\left(\beta_\pi + C \beta_\gamma\right) \pm {1 \over 2} \left(m \Omega_\text{env} \omega (1 \mp D) \over \alpha - D \omega^2 \right)\left((\beta_\pi - C \beta_\gamma)^2 \over \beta_\pi + C \beta_\gamma \right) + \mathcal{O}(\Omega^3).
\end{equation}

In the limit of weak coupling, the first term exhibits the same behaviour as the intrinsic asymmetry from second-order dynamical effects (as it arises from the convergent part of the series expansion), and decreases with increasing frequency as \(\Omega_\text{env} / \omega\); it is small for slow rotation. The second term, however, is potentially very large, as \(\alpha/\omega^2\) decreases rapidly with evolution, and we therefore expect it to dominate close to resonance. Thus, we have \edit1{an estimate of} the magnitude of the splitting asymmetry arising from near-resonance effects:
\begin{equation}
    \psi_\text{mix, two modes} \sim {1 \over 2} \left(m \Omega_\text{env} \omega \over \alpha\right)\left((\beta_\pi - C \beta_\gamma)^2 \over \beta_\pi + C \beta_\gamma \right).\label{eq:psi:mixbound1}
\end{equation}

Finally, we consider \(\epsilon\), the relative systematic error in measuring the rotational splittings induced by ignoring near-degeneracy effects, which we evaluate using \cref{eq:approxdev}. We find (setting \(D \to 0\) to simplify an otherwise very cumbersome expression) that
\begin{equation}
    \epsilon \sim  {m^2\Omega_\text{env}^2 \omega^2 \over 8 (\omega^2 \pm \alpha)^2} \left( -(C^2 \beta_\gamma^2 - 6 C \beta_\gamma \beta_\pi + \beta_\pi^2) \mp 2 {\omega^2 \over \alpha}\left(\beta_\pi - C \beta_\gamma\right)^2\right) + \mathcal{O}(\Omega^4).
\end{equation}
Once again, in the limit of weak coupling, the first term (which goes as \(\Omega_\text{env}^2/\omega^2\)) is small for slow rotation, while the second (which is proportional to \(1/\alpha\)) dominates close to resonance. Our \edit1{estimate of} this error then goes as
\begin{equation}
    \epsilon_\text{mix, two modes} \sim {m^2 \Omega_\text{env}^2 \over 4\alpha}\left(\beta_\pi - C \beta_\gamma\right)^2.\label{eq:eps:mixbound1}
\end{equation}

\hypertarget{matrix-elements}{%
\subsection{Matrix Elements}\label{matrix-elements}}

Our preceding discussion has ignored the off-diagonal elements of the matrix \(\mathbf{R}\) in some basis: either in the natural basis of mixed modes, or in the modified basis of the isolated \(\pi\) and \(\gamma\) modes. This property is required in order for the truncation of \cref{eq:perturb} to first order in \(\lambda\) to be a good approximation, which assumption in turn fundamentally underpins \cref{eq:kern}, and other methods built upon it. These two sets of basis functions are related to each other by some linear transformation \(\mathbf{C}\), as mixed-mode eigenfunctions may be expressed as linear combinations of \(\pi\) and \(\gamma\) modes. Thus, if \(\mathbf{R}^{\pi\gamma}\) is the representation of the rotation operator \(\mathcal{R}\) in the isolated \(\pi\) and \(\gamma\)-mode basis, then its corresponding representation in the natural mixed-mode basis is given as \(\mathbf{R}^\text{mixed} = \mathbf{C}^T \mathbf{R}^{\pi\gamma} \mathbf{C}\).

That we should obtain two quite different expressions for the multiplet asymmetry, when setting \(\mathbf{R}\) to be diagonal in each basis separately, indicates that the structure of \(\mathbf{C}\) in red giants is such that \(\mathbf{R}^\text{mixed}\) and \(\mathbf{R}^{\pi\gamma}\) cannot both be diagonal simultaneously. It is easy to demonstrate this explicitly. In these red giants, the density of \(\gamma\)-modes is much higher than that of the \(\pi\)-modes, so we expect to see many mixed modes close to resonance with any given \(\pi\)-mode. Let us consider two such modes whose \(m=0\) mode frequencies are both close to that of the same \(\pi\)-mode. We write their eigenfunctions as
\begin{equation}
    \bm{\xi}_k \sim c_{\pi,k}\bm{\xi}_\pi + \sum_j  c_{\gamma,kj}\bm{\xi}_{\gamma,j}.
\end{equation}
These coefficients may be related to the usual asymptotic mixing function \(\zeta_k\) as \(\zeta_k \sim \sum_j \left|c_{\gamma, kj}\right|^2\), while \(c_{\pi, k}^2 \sim 1 - \zeta_k\) (following \ob). By assumption, both of these modes are relatively p-mixed, so \(\zeta_k\) is small (and \(\sqrt{1 - \zeta_k}\) is close to 1). Supposing that \(\mathbf{R}^{\pi\gamma}\) is approximately diagonal, it follows that the off-diagonal matrix element of \(\mathbf{R}^\text{mixed}\) for these two modes is given as
\begin{equation}
\begin{aligned}
    R^\text{mixed}_{kl} &= \left<c_{\pi,k}\bm{\xi}_\pi + \sum_j  c_{\gamma,kj}\bm{\xi}_{\gamma,j}, \mathcal{R} \left(c_{\pi,l}\bm{\xi}_\pi + \sum_j  c_{\gamma,lj}\bm{\xi}_{\gamma,j}\right)\right> \\
    &= c_{\pi, k}c_{\pi, l} R^{\pi\gamma}_{\pi\pi} + \sum_j c_{\gamma, kj} c_{\gamma, lj} R^{\pi\gamma}_{\gamma, jj} + \text{(off-diagonal entries of $\mathbf{R}^{\pi\gamma}$)}.
\end{aligned}
\label{eq:offdiag1}
\end{equation}
Since the \(\gamma\)-modes in these red giants are far denser than the \(\pi\)-modes, we expect the \(\gamma\)-mode cross coupling not to contribute significantly, so
\begin{equation}
    R^\text{mixed}_{kl} \sim 2 m \sqrt{(1 - \zeta_k)(1 - \zeta_l)}\ \beta_\pi \Omega_\text{env} + \mathcal{O}\left(\sqrt{\zeta_k \zeta_l}\right). 
    \label{eq:offdiag2}
\end{equation}
In principle a converse argument can be made to show that \(\mathbf{R}^\text{mixed}\) being diagonal precludes \(\mathbf{R}^{\pi\gamma}\) from being so. However, \dheuv{} have already shown that off-diagonal entries in \(\mathbf{R}^\text{mixed}\) are required to reproduce the observed asymmetric quadrupole splitting in subgiants. In more evolved red giants, we expect at least two such dipole mixed modes with nontrivial off-diagonal elements to exist for every possible dipole \(\pi\) mode. The precise number of such mixed modes per p-mode that cannot be neglected will depend on how \(\zeta\) changes with frequency, which in turn depends on the structure of mode cavities and the strength of the coupling between them.

Let us now compare these expressions with those in earlier studies of asymmetric splitting. \dheuv{} provide explicit expressions for these terms in the two-mode, two-zone model. For two mixed modes in particular, the mixing coefficients in our construction must satisfy \((c_{\pi, 1}, c_{\gamma, 1}) = (c_{\gamma, 2}, -c_{\pi, 2})\) for orthogonality. Then \cref{eq:offdiag1} yields
\begin{equation}
    R_{12}^\text{mixed} = 2 m \sqrt{\zeta_1 (1-\zeta_1)} \left(\beta_\pi \Omega_\text{env} - \beta_\gamma \Omega_\text{core}\right),\label{eq:offdiag3}
\end{equation}
which reproduces the expression in \dheuv{} for the off-diagonal matrix element (their eq. B.7). Likewise, for the diagonal elements we recover
\begin{equation}
    R_{ii}^\text{mixed} = 2 m \left(\zeta_i \Omega_\text{core} \beta_\gamma + (1-\zeta_i) \Omega_\text{env} \beta_\pi\right),\label{eq:ondiag}
 \end{equation}
which ultimately yields \cref{eq:zetarot}. Since the resulting coupling matrices are identical, we conclude that the construction in \dheuv{} is equivalent to ours, and thus also assumes implicitly that the rotation matrix \(\mathbf{R}^{\pi\gamma}\) is diagonal in the isolated \(\pi/\gamma\) basis. Conversely, however, the constraint on the mixing coefficients required to produce \cref{eq:offdiag3} is only valid for two interacting modes. Thus, \cref{eq:offdiag3} fails in the r\'egime of many-\(\gamma\)-to-one-\(\pi\) mode coupling, which is the case for dipole mixed modes in red giants.

\citet{mosser_spin_2012,mosser_period_2015,mosser_period_2018} propose an alternative construction in this many-to-one r\'egime, in which the coupling fractions \(\zeta\) are themselves assumed to admit analytic continuation as a function of mixed-mode frequency, describing an infinitely dense forest of g-modes. Supposing this to be the case, their expressions for the rotational splitting (again in the case of two-zone differential rotation) reduce to requiring that the rotationally split mixed-mode frequencies satisfy
\begin{equation}
    \omega(\lambda) - \omega(0) = \left[ \int_{\omega(0)}^{\omega(0) + m\lambda \beta_\gamma \Omega_\text{core}}\zeta(\omega_g)\ \mathrm d \omega_g +  \int_{\omega(0)}^{\omega(0) + m \lambda\beta_\pi \Omega_\text{env}}\left(1- \zeta(\omega_p)\right)\ \mathrm d \omega_p \right],\label{eq:intzeta}
\end{equation}
in a self-consistent fashion, so that the application of Leibniz's theorem permits the leading order term, \(\mathrm{d}\omega\over\mathrm{d}\lambda\), to recover the diagonal matrix elements of \(\mathbf{R}^\text{mixed}\), \cref{eq:ondiag}. Since \(\zeta(\omega + \delta\omega_\text{rot}) \ne \zeta(\omega - \delta\omega_\text{rot})\) in general, this construction yields asymmetric splitting, and therefore implies the presence of non-zero rotation matrix elements off the diagonal. In order to compare their construction to that presented in this work, we will now derive expressions for these implied matrix elements.

Applying \cref{eq:approxasym}, we obtain that
\begin{equation}
    \psi \sim {\lambda \over 2} \left.\left( m \left(\left(\beta_\gamma \Omega_\text{core}\right)^2 - \left(\beta_\pi \Omega_\text{env}\right)^2\right) {\partial \zeta \over \partial \omega}\right) \right/ \left(\zeta\beta_\gamma \Omega_\text{core} + (1 - \zeta) \beta_\pi \Omega_\text{env}\right).
\end{equation}
We relate these derivatives to the off-diagonal matrix elements via \cref{eq:perturb}, yielding
\begin{equation}
    \sum_{j \ne i} {\left|R_{ij} \right|^2 \over \omega_{i,0}^2 - \omega_{j,0}^2} \sim 2 m^2\left[\left(\beta_\gamma \Omega_\text{core}\right)^2 - \left(\beta_\pi \Omega_\text{env}\right)^2\right] \left.{\partial \zeta \over \partial \omega^2}\right|_{\omega = \omega_i} + \mathcal{O}(\delta\omega_\text{rot}^2/\omega_i^2). \label{eq:zeta1}
\end{equation}
This does not in itself uniquely specify the matrix elements \(R_{ij}\). Ultimately, this reflects an implicit limitation of the \(\zeta\)-function construction: without further analytic continuation (e.g.~Ong \& Gehan in prep.), rotational coupling between modes via off-diagonal matrix elements is tacitly assumed to be strictly local, for infinitesimally separated modes in a continuum of possible mode frequencies. As we will see, this assumption is not consistent with the actual behaviour of the rotational coupling matrices returned from explicit numerical calculations. This also does not generalise to a discrete set of frequency eigenvalues in a well-defined manner. For the sake of argument, however, let us suppose that we may approximate the derivative in \cref{eq:zeta1} by some kind of finite difference scheme, where we choose coefficients \(h_{ij}\) so that
\begin{equation}
    \left.{\partial \zeta \over \partial \omega^2}\right|_{\omega = \omega_i} \sim \sum_{j \ne i} h_{ij}{\zeta_j - \zeta_i \over \omega^2_j - \omega^2_i}.
\end{equation}
Inserting this into \cref{eq:zeta1} then yields
\begin{equation}
    |R_{ij}|^2 \sim 2 m^2 h_{ij} (\zeta_i - \zeta_j) \left[\left(\beta_\gamma \Omega_\text{core}\right)^2 - \left(\beta_\pi \Omega_\text{env}\right)^2\right].
\end{equation}
Even were a canonical choice of \(h_{ij}\) to exist, expressions of this kind, irrespective of the precise choice of \(h_{ij}\), are clearly not consistent with the definition of the rotation matrix elements, \cref{eq:rotkernel}. In the two-zone model of differential rotation, \cref{eq:rotkernel} requires that \(R_{ij}\) must be proportional to some linear combination of \(\Omega_\text{core}\) and \(\Omega_\text{env}\), whereas this expression demands instead that its square be a linear combination of their squares. Accordingly, this construction is strictly speaking only valid in the limit of \(\Omega_\text{env} \to 0\), and cannot be used as a starting point for generalising the two-zone model of differential rotation in a manner consistent with the usual rotational kernels.

\begin{figure}
\centering
\includegraphics{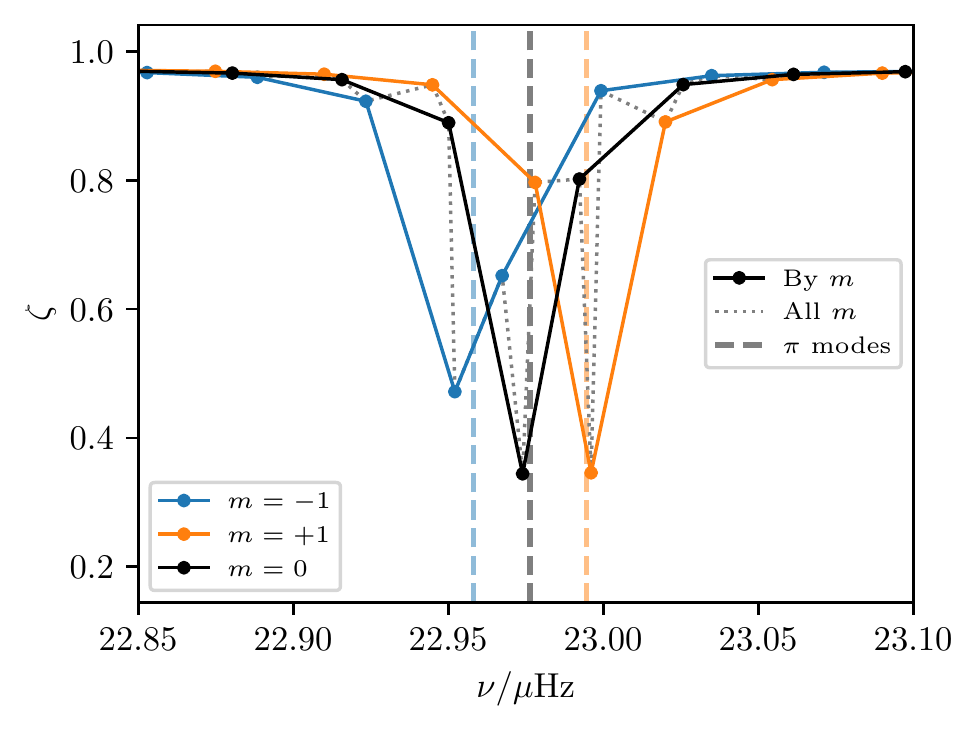}
\caption{Mixing functions \(\zeta\) for different azimuthal indices \(m\) (solid curves of different colours), computed using inertia ratios with respect to a red giant evolutionary model (Model 2 of the next section) with nontrivial envelope rotation. The dashed lines show the locations of the rotationally split \(\pi\) modes. The dotted lines connect the inertia ratios of all modes in increasing order of frequency; the resulting curve does not describe a mixing function \(\zeta\) of the kind used by the construction of \citet{mosser_period_2015} and subsequent works. \label{fig:zeta}}
\end{figure}

Another, less obvious, limitation of this construction is that the assumption of a single mixing function \(\zeta\) is not tenable even in the two-zone model. In the preceding discussion, we have assumed that the precise functional form of \(\zeta\) does not change, and is consistent with what would be obtained for a nonrotating star. In this two-zone model, however, changing \(\Omega_\text{env}\) changes the shape of \(\zeta\) itself, depending on whether prograde or retrograde modes are being considered (which we illustrate in \cref{fig:zeta}). The location of \edit1{each} local minimum of \(\zeta\) is determined by the location of the \(\pi\) mode of the appropriate \(m\) after accounting for its rotational perturbation, while the present construction of \citet{mosser_spin_2012,mosser_period_2018} etc. makes no allowances for this. Even if a modified formulation of this construction were to exist which resolves the issues above, it must \edit1{thus} also account for the splitting of the \(\pi\)-modes separately from that of the mixed modes in order to accurately describe mode splitting for nonzero \(\Omega_\text{env}\). \edit1{As such,} making reference to the \edit1{notional underlying basis of p- and g-modes, which we approximate with \(\pi\) and \(\gamma\) modes,} ultimately appears to be necessary even in the limit of infinitely dense \(\gamma\)-modes. \edit1{Corollarily, modes of different $m$ cannot be all straightened simultaneously on the same ``stretched'' echelle diagram \citep[as used in e.g.][]{gehan_core_2018}, in the presence of nontrivial envelope rotation.}

In summary, we have shown that our construction, further assuming that the off-diagonal elements of \(\mathbf{R}^{\pi\gamma}\) may be neglected, is fully equivalent to the formulation of \dheuv{} when restricted to the two-multiplet two-zone problem, \edit1{and} is also immediately generalisable to the extended case of many-mode coupling. We have moreover demonstrated that the implied off-diagonal matrix elements from the construction of \citet{mosser_spin_2012} etc. are not only not uniquely defined, but are also in any case inconsistent with the linear property of the matrix elements of the perturbing rotation operator, \cref{eq:rotkernel}. Any similar techniques based on the approximate asymptotic mode-bumping function \(\zeta\) also cannot correctly describe the splitting of mixed modes for nonzero \(\Omega_\text{env}\), unless the shape of \(\zeta\) should itself also be made to depend parametrically on the splitting of the underlying \(\pi\) modes.

\hypertarget{numerical-results}{%
\section{Numerical Results}\label{numerical-results}}

\label{sec:numerics}

Thus far, we have not supplied any a priori justification for neglecting the off-diagonal elements \(R_{ij}^{\pi\gamma}\). To the extent that they may be neglected for pure p- and g-modes \edit1{for stars in other stages of evolution}, it seems reasonable that the off-diagonal elements in the \(\pi\) and \(\gamma\) subspaces may be ignored, as they separately behave like pure p- and g-modes. Moreover, since the \(\pi\)-modes are formally evanescent where the \(\gamma\)-modes may propagate (and vice versa), we should also expect their overlap terms in \cref{eq:rotkernel} to be negligible. In order to make more concrete statements about how significant these off-diagonal elements might actually be, however, we will have no choice but to resort to explicit numerical calculations. To illustrate the properties of the above expressions, we evaluate them with respect to stellar structures specified by evolutionary models computed using \mesa{} r12778 \citep{mesa_paper_1,mesa_paper_2,mesa_paper_3,mesa_paper_4,mesa_paper_5}. We consider models along these evolutionary tracks from the onset of mode mixing (starting where \(\numax\Delta\Pi_1 < 1\)) to the tip of the red giant branch. We evaluate \cref{eq:rotkernel} and \cref{eq:centkernel} using frequency eigenvalues and mode eigenfunctions computed using \gyre{} \citep{townsend_gyre_2013}. These are evaluated both with respect to the usual basis of mixed modes, as well as with respect to the \(\pi\) and \(\gamma\) modes and their coupling matrices, which we computed, also using \gyre, according to the prescription of \citet{ong_semianalytic_2020}.

\hypertarget{rotational-matrices}{%
\subsection{Rotational Matrices}\label{rotational-matrices}}

\begin{figure*}[htbp]
    \centering
    \includegraphics[width=\textwidth]{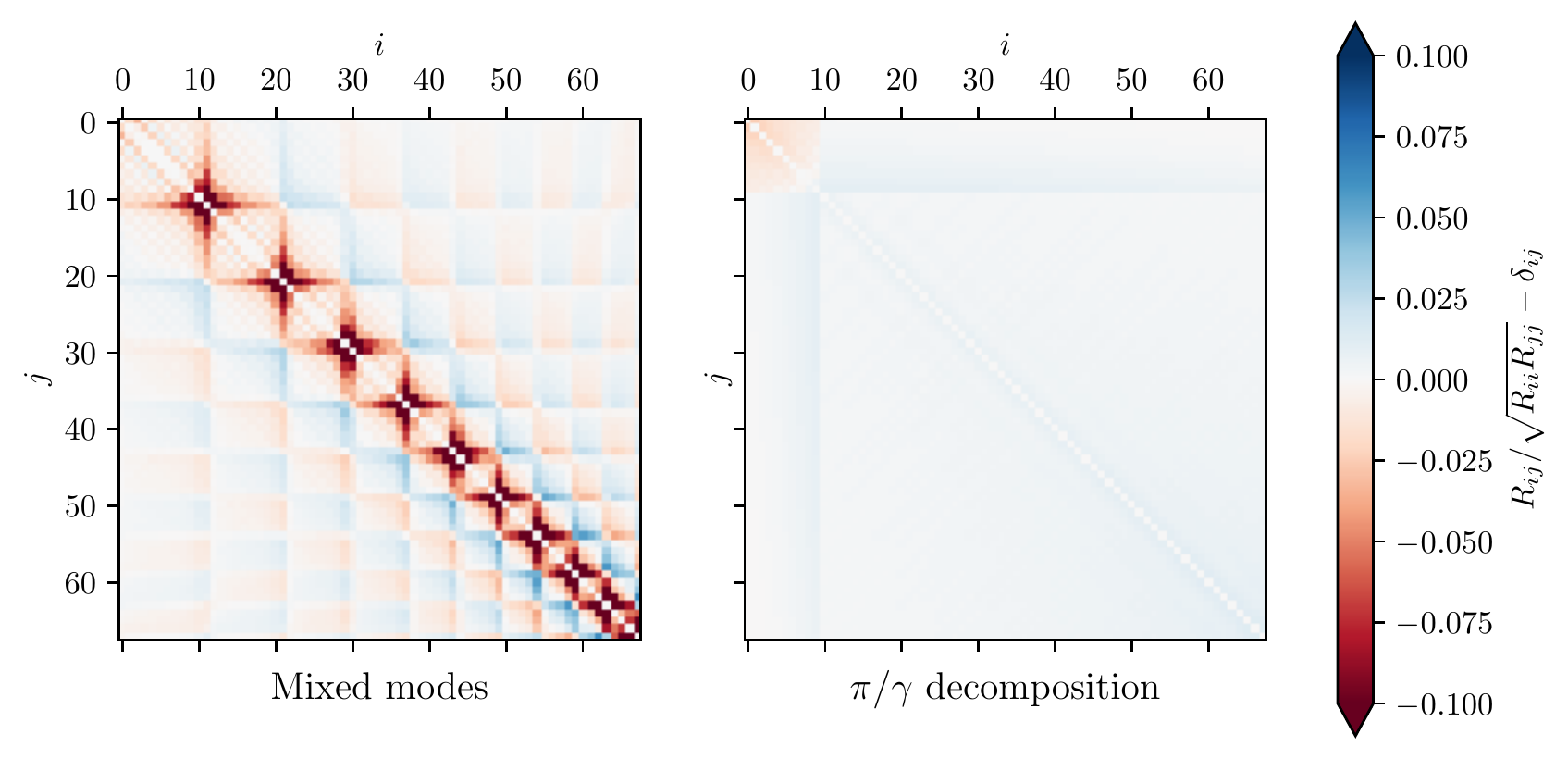}
    \caption{Scaled off-diagonal elements of the matrix elements $R_{ij}$ with respect to two different sets of basis functions for dipole modes. Matrix elements in the natural basis of mixed modes are shown in the left panel, and those in the isolated basis of $\pi$ and $\gamma$ modes are shown in the right panel. Note that both panels use the same colour scale, which is significantly saturated in the left panel in order to show structure in the right panel.}
    \label{fig:off_diagonal}
\end{figure*}

For illustrative purposes, we first apply this numerical construction to a model of a young red giant (\(\Dnu=17.5\ \mu\)Hz) with mass \(1.4\ M_\sun\), of solar composition and mixing length, using a notional value of \(\Omega_\text{core} / 2\pi = 0.5 \mu\)Hz and \(\Omega_\text{core}/\Omega_\text{envelope}\approx 10\), as in e.g. \citet{gehan_core_2018} and \citet{bugnet_magnetic_2021}. Since we rely heavily on this model for demonstrative purposes, we will refer to it hereafter as ``Model 1''. It is also convenient to describe the configuration of mixed modes using the number of dipole g-modes per p-mode at \numax, \(\mathcal{N}_1 = \Delta\nu / \nu^2 \Delta\Pi_1\). For Model 1, we have \(\mathcal{N}_1 \sim 5\). We show the structure of the off-diagonal elements of its dipole-mode rotation matrices (\(m = l = 1\)) in \cref{fig:off_diagonal}, scaled by its diagonal elements in order to produce dimensionless quantities that can be compared between the two sets of basis functions. In particular, we see that while \(\mathbf{R}^\text{mixed}\) plainly exhibits very significant structure off the diagonal, \(\mathbf{R}^{\pi\gamma}\) does not: the off-diagonal elements of the former are an order of magnitude larger than the latter.

\begin{figure}
\centering
\includegraphics{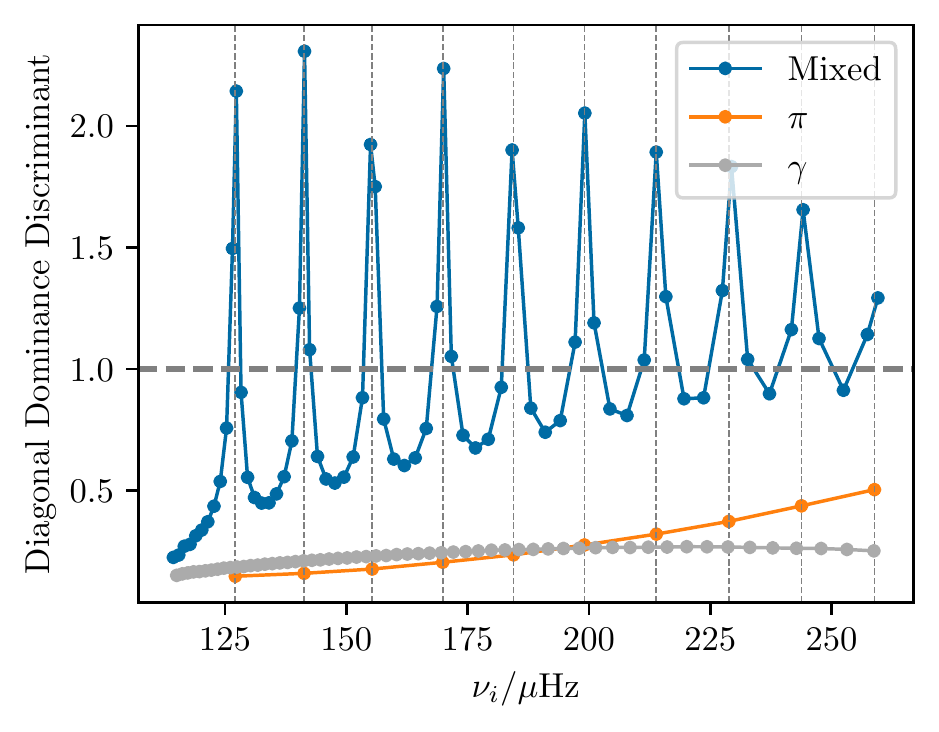}
\caption{Diagonal-dominance discriminant for the mixed, \(\pi\), and \(\gamma\) dipole-mode rotational matrices (including cross-coupling terms) shown in \cref{fig:off_diagonal}. Modes for which this discriminant exceeds unity, shown with the horizontal dashed line, are most poorly approximated by a diagonal rotation matrix. The \(\pi\)-mode frequencies are shown with the vertical dashed lines. \label{fig:dominance}}
\end{figure}

\cref{eq:offdiag2} predicts that the off-diagonal elements in these red giants are dominated by terms proportional to \(\sqrt{(1 - \zeta_{i})(1 - \zeta_{j})}\), and therefore are largest for pairs of p-dominated mixed modes. It is precisely this that gives rise to the plus-shaped structure in the left panel of \cref{fig:off_diagonal}. For these modes specifically, the rotation matrix is most poorly approximated as being diagonal. To better quantify this statement, we define a preconditioned diagonal dominance discriminant,
\begin{equation}
    S_i = \sum_{j \ne i} {|R_{ij}| \over \sqrt{|R_{ii}R_{jj}|}},
\end{equation}
such that the rotation matrix is diagonally dominant if \(S_i < 1\) for all modes \(i\). We show these values for the mixed modes, \(\pi\) modes, and \(\gamma\) modes in \cref{fig:dominance}. To permit a fair comparison, we include the \(\pi\)-\(\gamma\) cross-coupling terms of \(\mathbf{R}^{\pi\gamma}\) when evaluating \(S_i\) for the \(\pi\) and \(\gamma\) modes. Nonetheless, we see that \(\mathbf{R}^{\pi\gamma}\) is indeed diagonally dominant, while the modes for which off-diagonal elements of \(\mathbf{R}^\text{mixed}\) can least be neglected are precisely the most p-dominated mixed modes, whose frequencies are closest to those of the underlying \(\pi\) modes. \edit1{Accordingly, neglecting the off-diagonal elements of \(\mathbf{R}^{\pi\gamma}\) is a far better approximation than doing so for \(\mathbf{R}^\text{mixed}\).}

\begin{figure}
\centering
\includegraphics{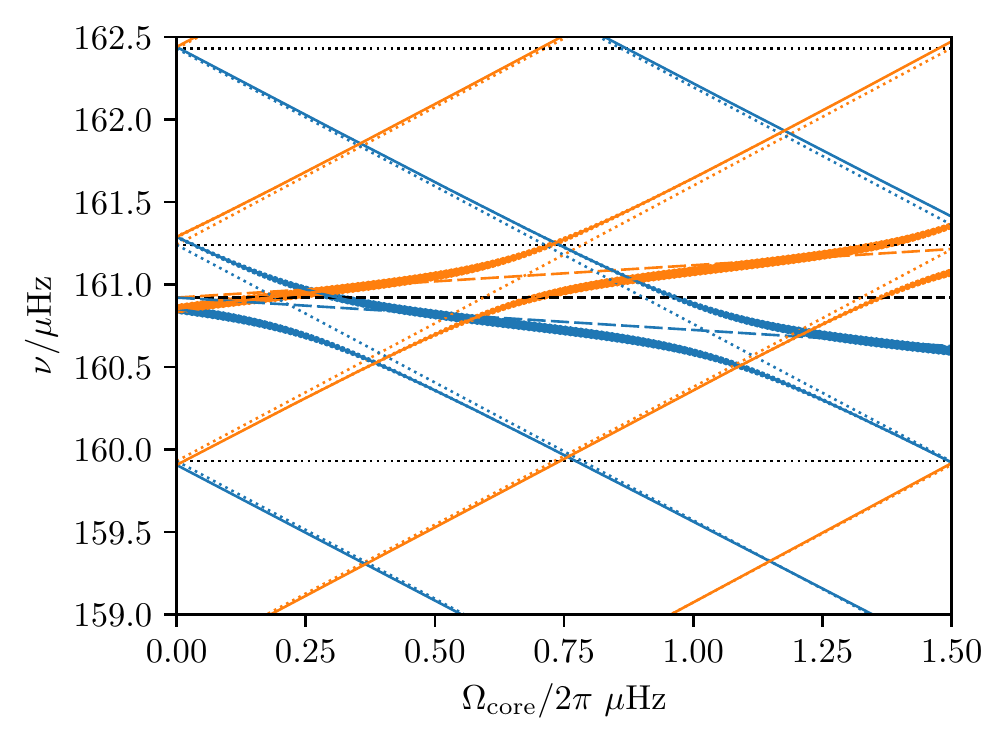}
\caption{Mixed-mode rotational splittings as a function of \(\Omega\), for quadrupole modes in the same red giant model used for \cref{fig:off_diagonal}. Retrograde sectoral modes are shown in blue, prograde sectoral modes in orange, and zonal modes in grey. Mixed modes are shown with solid lines, while the uncoupled p- and g-modes (and their splittings) are shown with the dashed and dotted lines in the background, respectively. Note the avoided crossings in the solid lines near where dotted and dashed lines of the same colour cross. \label{fig:avoidedcrossings}}
\end{figure}

These off-diagonal matrix elements \edit1{of \(\mathbf{R}^\text{mixed}\)} introduce coupling between \edit1{different mixed} modes, as described in \dheuv, resulting in avoided crossings between modes of the same \(m\) in different multiplets. To build intuition for this, we show the frequencies of the sectoral quadrupole modes from the same young red giant model as before (where the phenomenon is more visually evident), as a function of the core rotation rate, in \cref{fig:avoidedcrossings}, maintaining a constant value of the core-envelope rotational contrast \(C = 10\) throughout. We see that our construction reproduces the qualitative features also observed from nonperturbative calculations for mixed-mode rotational splittings, e.g.~as performed in \citet{ouazzani_rotation_2013} and \dheuv. In particular, the mixed-mode frequencies, represented with the solid curves, generate independent families of avoided crossings for each azimuthal order \(m\), shown with different colours. We additionally represent variations in the mixing fraction \(\zeta\) using the stroke thickness, with the g-dominated mixed modes shown using thin lines, and p-dominated ones with thick lines. Away from resonance, the splittings of the g-dominated mixed modes are well-described by those of the pure \(\gamma\)-modes, shown with the dotted lines; likewise the splittings of the p-dominated mixed modes approximate those of the pure \(\pi\)-modes, shown with the dashed lines. It is when these avoided crossings occur --- i.e.,~when these families of straight lines cross --- that we get deviations from symmetric rotational splitting. We stress that these avoided crossings are parameterised with respect to \(\Omega\), rather than evolution as ordinarily considered.

\edit2{The often-used linear expression for mixed modes, \cref{eq:zetarot}, predicts that rotational splittings for mixed modes should also yield straight lines on \cref{fig:avoidedcrossings} at an intermediate angle between the pure p- and g-modes, to leading order in perturbation theory. We can now see that it is the curvature of the near-degenerate mixed-mode avoided crossings which require second- and higher-order terms involving the off-diagonal matrix elements to fully describe. By contrast, these rotational effects are much simpler to conceptualise with access to the pure p- and g-modes and the coupling between them: these isolated pure modes can be seen to trace out the straight lines in \cref{fig:avoidedcrossings}, and may then be recombined to yield mixed modes by effecting the coupling between the two mode cavities. Importantly, this mode coupling does not have any dynamical dependence on the rotation rate. Thus, a linear treatment of rotational effects remains applicable to the pure p- and g-modes even when it does not for mixed modes}.

\hypertarget{estimates-of-asymmetric-splitting}{%
\subsection{Estimates of Asymmetric Splitting}\label{estimates-of-asymmetric-splitting}}

\begin{figure}[htbp]
    \centering
    \annotate{\includegraphics[width=.485\textwidth, trim=.25cm .25cm .25cm .25cm,clip]{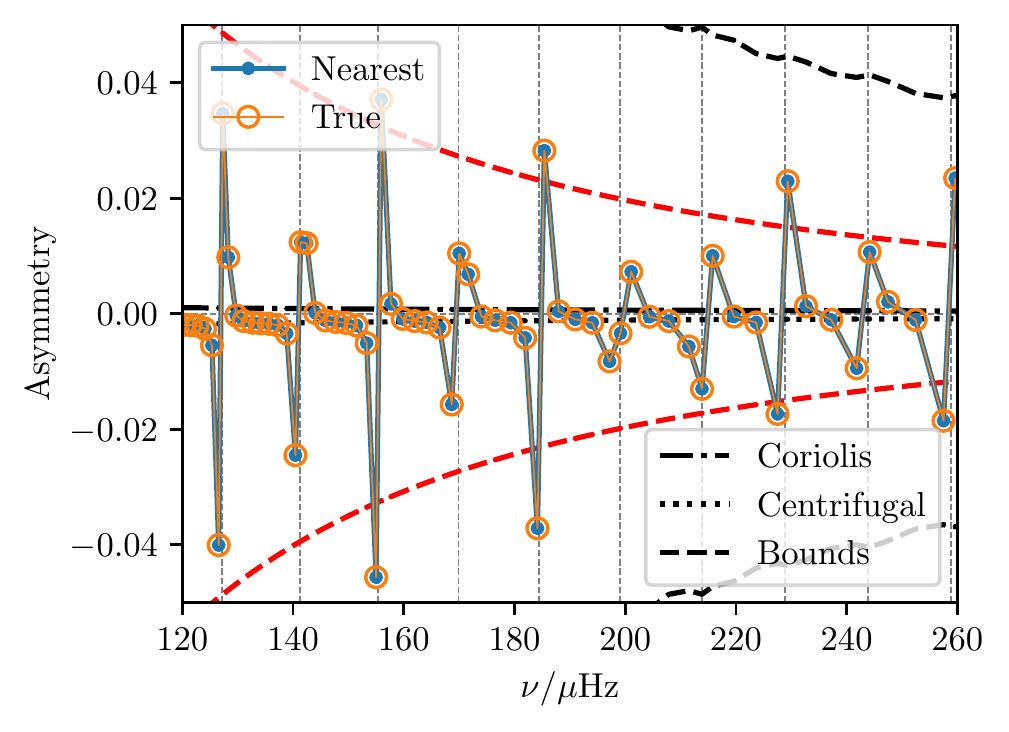}}{\node[left] at (.95, .85){\textbf{(a)}: $l=1$};}
    \annotate{\includegraphics[width=.485\textwidth, trim=.25cm .25cm .25cm .25cm,clip]{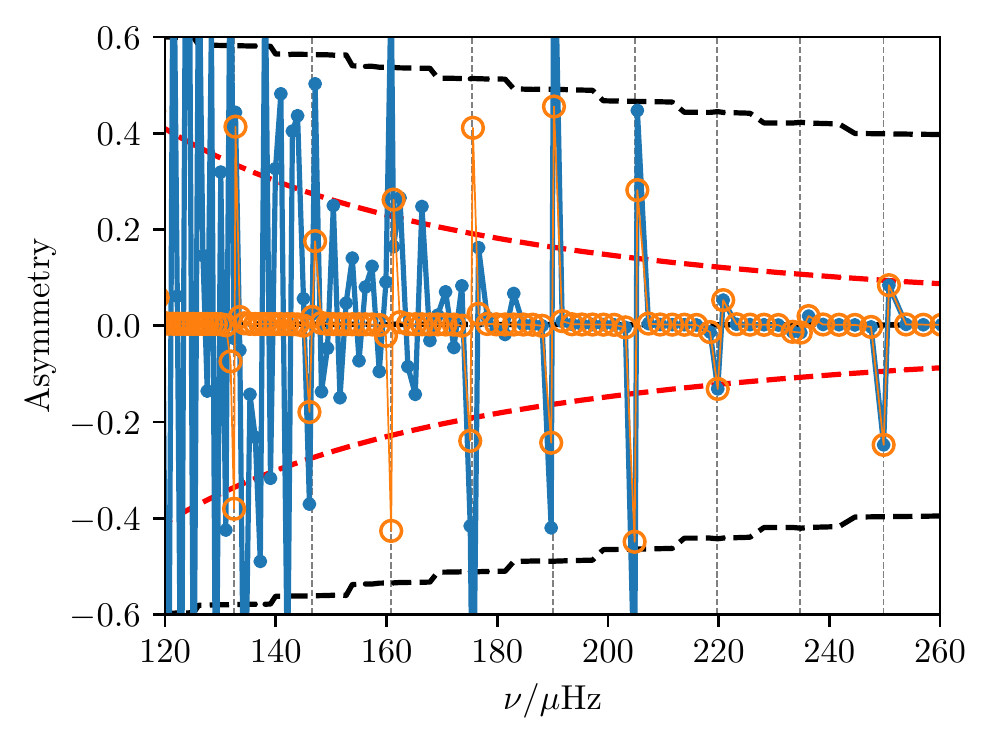}}{\node[left] at (.95, .92){\textbf{(b)}: $l=2$};}
    \caption{Asymmetry parameters $\psi$ for \textbf{(a)} dipole and \textbf{(b)} quadrupole modes, for a MESA evolutionary model near the base of the red giant branch (see text for complete description). Markers and lines connecting them show the asymmetry in the mode splitting computed using \cref{eq:asym}, both with respect to the actual multiplets (orange open circles and lines), as well as for multiplets comprised of the two nearest sectoral modes to each zonal mode (blue markers and lines). Other curves show various estimates for asymmetric multiplet splitting: we show predictions from only second-order Coriolis effects (dot-dashed black lines), from second-order rotational effects including the centrifugal force (dotted lines), and the upper bound on asymmetric splitting from near-degeneracy effects (\cref{eq:psi:mixbound1}, black dashed lines). A modified expression for near-degeneracy effects accounting for the dense g-mode spectrum, \cref{eq:psi:mixbound2}, is shown with the red dashed lines. The locations of the $\pi$-modes are shown with vertical dashed lines, to guide the eye.}
    \label{fig:asym1}
\end{figure}

Let us now examine the effects of these avoided crossings on the sizes and asymmetries of the mixed-mode splittings themselves. We illustrate this by computing the asymmetry parameter \(\psi\) directly using \cref{eq:asym} with respect to only the sectoral and zonal mode frequencies of Model 1, as in \dheuv{}, again using a core rotation rate of \(0.5\ \mu\)Hz and a core-envelope contrast of 10. At lower frequencies, \(\Omega_\text{core}\) is comparable to the frequency difference between adjacent \(g\)-like mixed modes. This may cause confusion in identifying which modes should constitute the multiplet of a given \(l\) and \(n\). In principle, the zonal and sectoral modes should nonetheless remain distinguishable by way of their relative amplitudes \citep[with some inclination dependence, as in e.g.][]{gizon_inclination_2003}. Accordingly, we compute \(\psi\) in two ways. Since the quantum numbers of all modes are known a priori, we are able to present the ``true'' asymmetry parameter, shown with orange lines and open circles. Additionally, we show, with the blue markers and lines, a naive ``nearest-neighbour'' construction, which assumes that an unsuspecting observer will have little choice but to assign the nearest two sectoral modes to each zonal mode. We show the signed asymmetry from both constructions in \cref{fig:asym1}, for both dipole and quadrupole mode frequencies. Superimposed on these values, we also show three different \edit1{estimates} of the asymmetry in the multiplet splitting: \edit1{that arising} from the second-order Coriolis effect alone (dot-dashed line), that from both the second-order Coriolis and centrifugal forces combined (\cref{eq:bareasym}, dotted lines), and the \edit1{limiting value} from near-degeneracy effects (\cref{eq:psi:mixbound1}, black dashed lines), \edit1{which can be seen to serve as an upper bound}. To guide the eye, we show the locations of the \(\pi\) modes (and thus the most p-dominated mixed modes) with the vertical dashed lines.

For dipole modes (panel a), we see that the size of the splitting is small enough that there is no confusion between the two constructions. For the most g-dominated mixed modes, the splitting asymmetry is also small, and well-described purely by higher-order dynamical effects. While we have considered only the second-order dynamical effects of rotation here, the higher-order structural effects of rotation (i.e.~changes to mode frequencies owing to changes to stellar structure caused by rotation) enter into the wave equation to the same order in \(\Omega\); their effects will therefore be likewise small. In the vicinity of the more p-dominated mixed modes, however, the resonant asymmetries induced by avoided crossings clearly dominate this intrinsic dynamical asymmetry. For the quadrupole modes, we see in panel b that the width of the rotational splitting becomes comparable to the spacing of the g-modes, rendering them susceptible to misidentification with the naive nearest-neighbour approach, particularly at low frequencies. Again, asymmetric splitting arises predominantly resonantly in the most p-dominated mixed modes from avoided crossings involving the underlying \(\pi\)-modes, and by far exceeds the small amounts generated by second-order dynamical effects. Moreover, the coupling strength between the p- and g-mode cavities is much weaker for quadrupole modes than for dipole modes; this coupling strength appears in the denominator of \cref{eq:psi:mixbound1,eq:eps:mixbound1}. Accordingly, we expect that the potential asymmetric splitting from near-resonance effects to be much larger for quadrupole modes than for dipole modes, which is indeed seen to be the case.

\begin{figure}[htbp]
	\centering
	\includegraphics{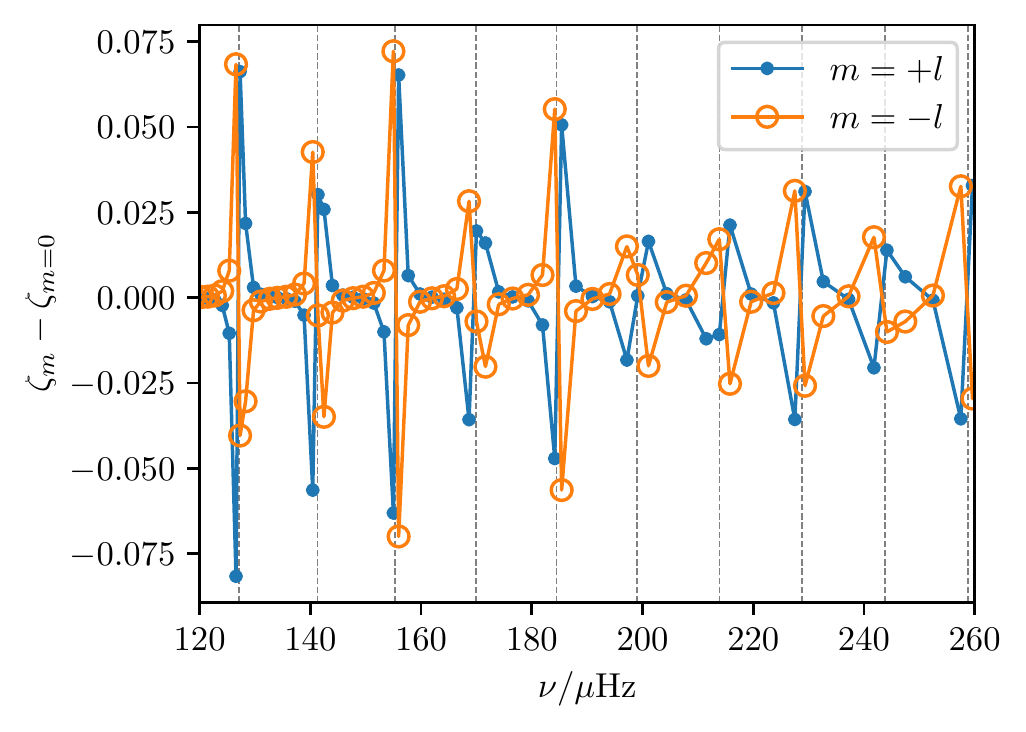}
	\caption{Differences between the mixing fraction $\zeta_m$ of multiplet components of different azimuthal orders $m$ with that of the $m=0$ multiplet component, $\zeta_0$, shown as a function of the nonrotating mode frequency for the dipole modes of model 1 (cf. \cref{fig:asym1}a). Vertical dashed lines indicate the locations of the underlying $\pi$-modes.}
	\label{fig:dzeta}
\end{figure}

We note that this phenomenology, with the unsigned asymmetry taking maximal values for the most p-dominated modes, is the converse of that demonstrated with less evolved subgiants in \dheuv. In that work, asymmetric splitting was shown to be associated with avoided crossings producing g-dominated mixed modes, amidst otherwise symmetrically-split p-modes. However, the intrinsically asymmetric splitting in both of these cases arises through essentially the same mechanism, which is perhaps most easily visualised by considering the mixing fractions $\zeta$ of each of the multiplet components separately. In the specific case of only core rotation in red giants, we may combine \cref{eq:approxasym} with the defining property of $\zeta$ in the continuum limit of g-mode frequencies, $\zeta \sim \partial \omega / \partial \omega_g$ (cf. \cref{eq:intzeta}), to obtain roughly that
\begin{equation}
\zeta_m - \zeta_0 \sim m \beta_\text{core} \Omega_\text{rot} {\partial \zeta \over \partial \omega_g} \sim (m \beta_\text{core} \Omega_\text{rot})^{-1} {\partial^2 \omega \over \partial \lambda^2} \sim {2 \over m} \psi. \label{eq:dzeta}
\end{equation}
While this precise chain of reasoning is not generally valid (following our earlier discussion in \autoref{matrix-elements}), it nonetheless suffices to motivate a heuristic diagnostic for determining where asymmetric rotational splittings should be expected to emerge: it arises where the mixing fractions $\zeta_m$ are significantly different between components of the same rotational multiplet. For illustration, we show these quantities in \cref{fig:dzeta} for the dipole modes of Model 1; despite the roughness of the assumptions going into \cref{eq:dzeta}, we can see that it produces good visual agreement with the actual intrinsic asymmetries shown in \cref{fig:asym1}a.

We now turn our attention to analytic limits on the asymmetry and systematic error. In both of its panels, \cref{eq:psi:mixbound1} can be seen to serve only as a loose upper bound, rather than being representatively descriptive of the typical resonant asymmetries in the neighbourhoods of these avoided crossings. Ultimately, this is because \cref{eq:psi:mixbound1} is derived for the specific case of exactly one \(\pi\) mode being fortuitously in resonance with exactly one \(\gamma\) mode, where both sets of modes are assumed to be otherwise sparse. In more evolved red giants, however, the forest of \(\gamma\) modes near each \(\pi\) mode becomes increasingly dense, spaced out at a local repetition rate of \(\delta\omega_\gamma = 2 \pi \nu^2 \Delta\Pi\), where \(\Delta\Pi\) is the characteristic period spacing of the \(\gamma\)-modes. Consequently, there will always be at least one \(\gamma\) mode at most \(\delta\omega_\gamma\) away from every \(\pi\) mode: the resonance factors in the second-order term of \cref{eq:expand} are then bounded from below by terms inversely proportional to \(\delta \omega_\gamma^2 \sim 2\omega\ \delta\omega_\gamma\). Compared to this, \citet{ong_surface_1} show that conversely the distance between adjacent mixed-mode eigenvalues is bounded from below as \(\delta \omega^2 > 2 \alpha\) in the two-mode system, which is why \cref{eq:psi:mixbound1} is an upper bound. Accordingly, to match the resonance-dominated terms of \cref{eq:expand}, we can modify \cref{eq:psi:mixbound1} by replacing \(\alpha\) in the denominator with \(\omega\ \delta\omega_\gamma\), to yield
\begin{equation}
    \psi_\text{mix} \sim {1 \over 2} \left(m \Omega_\text{env} \over 2\pi \nu^2 \Delta\Pi\right)\left((\beta_\pi - C \beta_\gamma)^2 \over \beta_\pi + C \beta_\gamma \right). \label{eq:psi:mixbound2}
\end{equation}
We show these values with the red dashed lines in \cref{fig:asym1}. For both sets of modes, \cref{eq:psi:mixbound2} is more representative of the near-degeneracy asymmetric splitting than the upper bound from \cref{eq:psi:mixbound1}.

\begin{figure}
\centering
\includegraphics{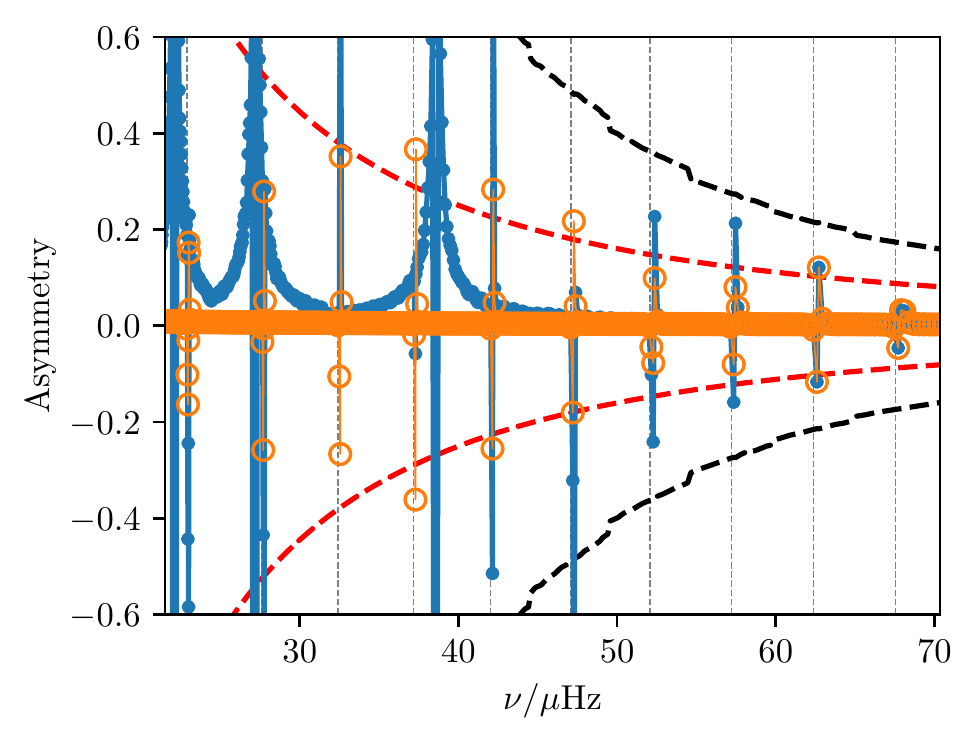}
\caption{The same quantities as shown in \cref{fig:asym1}, computed with respect to dipole modes in a more evolved red giant model (see text for a complete description).\label{fig:asym2}}
\end{figure}

These expressions appear to hold well more generally. We show in \cref{fig:asym2} the same quantities when repeating this exercise with dipole modes in a significantly more evolved red giant model (\(\Delta\nu = 5\ \mu\)Hz, \(\mathcal{N}_1 \sim 30\)) along the same evolutionary track as Model 1, which has evolved to the RGB luminosity bump (we will refer to this as Model 2). We perform calculations with the same rotational rates as above, although in reality we should expect the core-envelope rotational contrast to intensify over the course of stellar evolution. Nonetheless, this calculation still serves for our illustrative purposes. As \citet{gehan_core_2018} observe, correct identification of rotational multiplet components becomes increasingly difficult for more evolved stars. This is bourne out in our calculations by the clear disagreement between asymmetries calculated with the true mode identification in hand, compared to with the naive nearest-neighbour construction. At the same time, the coupling strengths for mixed modes also decrease rapidly over the course of stellar evolution \citep{ong_surface_1}, causing even the true splitting asymmetry here to be an order of magnitude larger than that of the dipole modes in Model 1. Again, we see that \cref{eq:psi:mixbound1} serves as a loose upper bound on the splitting asymmetry, while \cref{eq:psi:mixbound2} yields more representative predictions. We will thus use \cref{eq:psi:mixbound2} for our discussion in the subsequent sections.

\begin{figure}[htbp]
    \centering
    \annotate{\includegraphics[width=.485\textwidth, trim=.25cm .25cm .25cm .25cm,clip]{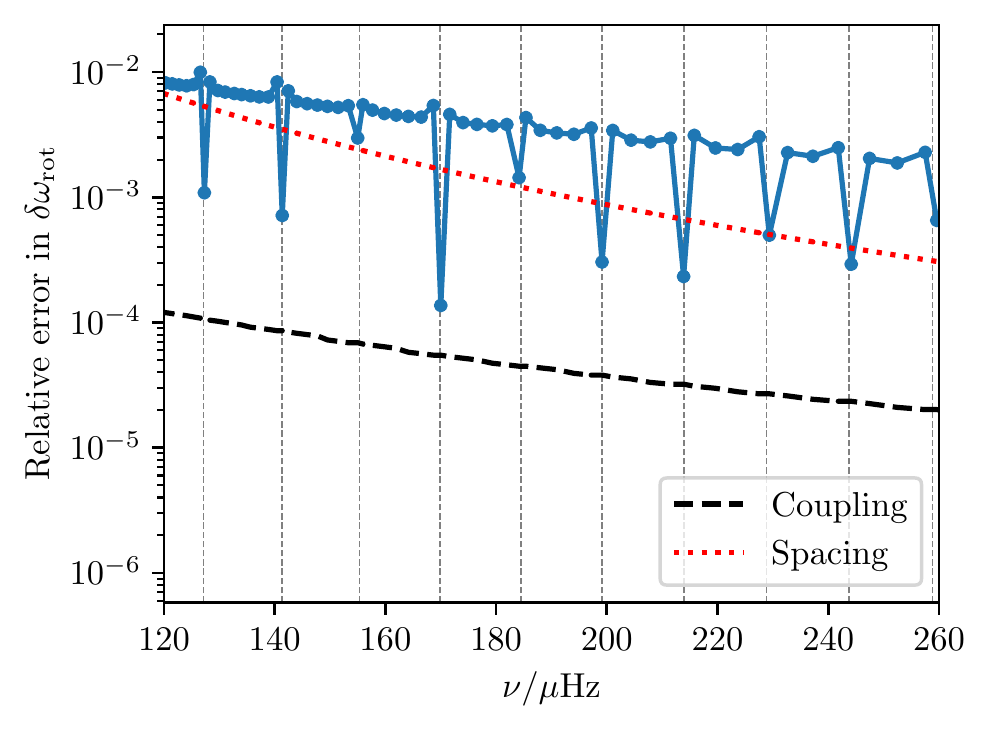}}{\node[left] at (.95, .95){\textbf{(a)}};}
    \annotate{\includegraphics[width=.485\textwidth, trim=.25cm .25cm .25cm .25cm,clip]{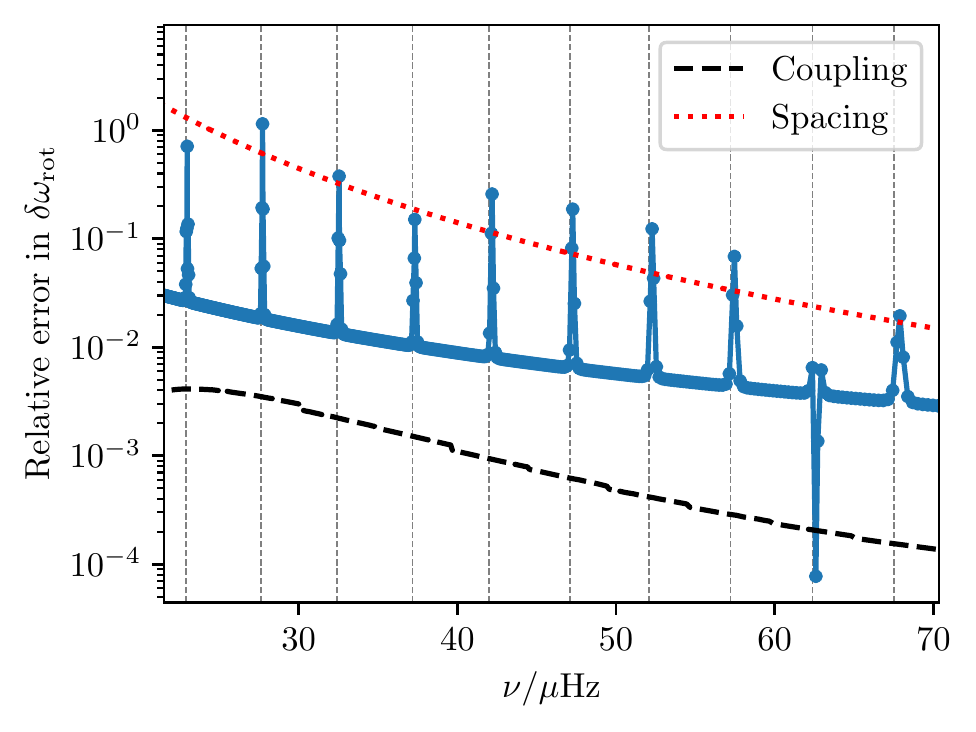}}{\node[left] at (.25, .2){\textbf{(b)}};}
    \caption{Relative error $\epsilon$ in the rotational width $\delta\omega_\text{rot}$ for dipole modes of MESA evolutionary models near \textbf{(a)} the base of the red giant branch, and \textbf{(b)} the luminosity bump (see text for complete description). Blue markers show the relative systematic errors in the rotational width compared to predictions using the first-order expressions, which are given in the two-zone model by \cref{eq:zetarot}. We also show expressions for this error owing to near-degeneracy effects as predicted from two-mode coupling alone (black dashed lines, \cref{eq:eps:mixbound1}), and accounting for the dense forest of g-modes (red dotted lines, \cref{eq:eps:mixbound2}). Note that the g-dominated modes have opposite signed error to the p-dominated modes.}
    \label{fig:err}
\end{figure}

Also of concern, aside from asymmetry per se, is the possible systematic errors induced in interpreting measurements of the rotational splitting as rotation rates. When near-degeneracy effects dominate, the first-order expressions in the two-zone model, \cref{eq:zetarot}, may cease to provide a good approximation to the true widths of the rotational splittings. We show the absolute relative systematic errors in these measurements, \cref{eq:dev}, in \cref{fig:err}, for dipole modes from both Models 1 and 2. Note that the g-dominated modes exhibit systematic deviations from the first-order expression as well; these are of opposite sign to those shown by the p-dominated modes. We also show the estimates for this relative error in the two-mode coupling scenario, \cref{eq:eps:mixbound1}, with the black dashed lines. It is evident that considerations from two-mode coupling significantly underpredict the systematic error in the rotational splitting widths relative to the first-order expression.

Again, we refer to the perturbative expansion, \cref{eq:expand}, for guidance as to how \cref{eq:eps:mixbound1} may be modified to suit a dense forest of g-modes. In the limit of degeneracy-dominated effects, the third-order term in the expansion is dominated by terms of the form
\begin{equation}
    \omega_{i,3} \sim \lambda^3  {\omega_{i,0}^2 \over 2}\sum_{j \ne i}\sum_{k \ne i}{R_{ij} R_{jk} R_{ki} \over \left(\omega^2_{i,0} - \omega^2_{j,0}\right)\left(\omega^2_{i,0} - \omega^2_{k,0}\right)}.
\end{equation}
Accordingly, in line with the above considerations, we propose a modified expression,
\begin{equation}
    \epsilon_\text{mix} \sim {1 \over 4}\left(m \Omega_\text{env} \over 2 \pi \nu^2 \Delta\Pi\right)^2\left(\beta_\pi - C \beta_\gamma\right)^2,\label{eq:eps:mixbound2}
\end{equation}
which scales appropriately with \(\Omega\) and \(\Delta\Pi\), and also retains the desirable property that this systematic error should vanish should the core-envelope rotational contrast be fortuitously equal to \(\beta_\pi / \beta_\gamma\). We show this expression with the red dotted lines in \cref{fig:err}; again we find that it is generally representative of the systematic error in the vicinity of the \(\pi\) modes.

\hypertarget{other-sources-of-asymmetric-splitting}{%
\subsection{Interactions with other sources of asymmetric splitting}\label{other-sources-of-asymmetric-splitting}}

\label{sec:mag}

In addition to the structural and dynamical effects that we have considered above, \edit1{frequency shifts depending on $m$} can also result from further violations of spherical symmetry, such as would arise from latitudinal differential rotation or large-scale magnetic fields. \edit1{As the former does not affect the asymmetry parameter $\psi$ \citep{aertsbook}, we restrict our attention to the latter.} \citet{bugnet_magnetic_2021} consider the effects of an axisymmetric magnetic field localised to the radiative core of a red giant on the rotationally-split mixed-mode frequencies. Since the magnetic field is localised to the core, by assumption its effects on the g-dominated mixed modes is much larger than on the p-dominated ones. They then consider the interposition of magnetic and rotational effects to first order, where frequency perturbations are strictly additive. With respect to this treatment, they find that magnetic fields also induce much larger rotational asymmetries for g-dominated mixed modes than for p-dominated mixed modes. However, as we have seen above, near-degeneracy effects systematically induce asymmetric splitting in p-dominated mixed modes, and these arise from nonlinear effects of mode coupling. Here, we examine how such nonlinearities might affect estimates for the asymmetric splitting arising from magnetic effects as well.

We assume that the star hosts a stable buried magnetic field in its radiative interior, resulting from the stabilization of past dynamo fields \citep{Arlt2013, Emeriau-Viard2017, Villebrun2019}. Purely toroidal and purely poloidal magnetic configurations are known to be unstable inside radiative interiors \citep[e.g.][]{Tayler1973, Markey1973, Braithwaite2006, Braithwaite2007a}, but the stability of mixed configurations with both poloidal and toroidal components has been demonstrated in \citet{Tayler1980}. We therefore use such a mixed poloidal and toroidal magnetic field configuration expressed analytically by 
\cite{Duez2010} to model a stable fossil field inside the radiative interior of the star, with its axis of symmetry aligned with the rotation axis of the star, as done in \citet{bugnet_magnetic_2021}. We ignore any dynamo action in the convective envelope, as the resulting magnetic field amplitude would be too low for its effect on mixed mode frequencies to be detectable \citep[e.g.][for magnetic cycle amplitudes at the surface of the Sun and red giants]{Perri2020, Privitera2016}. We set the maximum amplitude of the field at $B_0=1\ \mathrm{MG}$ for Model 1, as might be typical inside red giants' radiative interiors \citep{Cantiello2016} from the conservation of the magnetic flux inside the radiative region since the last convective-core event \citep[see][for more details about fossil fields conservation]{bugnet_magnetic_2021}. This field amplitude is small enough so that a first-order perturbation study can be applied, as in \citet{bugnet_magnetic_2021}.

We accommodate magnetic effects in our construction perturbatively by modifying the rotational QHEP to include a further operator describing the action of magnetic fields, as
\begin{equation}
    \left(\omega^2 \mathcal{D} + \lambda \omega \mathcal{R} + \mathcal{L} + \lambda^2\mathcal{V} + \kappa \mathcal{M} \right)\bm{\xi} = 0, \label{eq:qhep:mag}
\end{equation}
where \(\kappa \in [0, 1]\) is a perturbative expansion parameter (of the same kind as \(\lambda\)), and the elements of the corresponding matrix representation \(\mathbf{M}\) of the operator \(\mathcal{M}\) are given as
\begin{equation}
    M_{ij} = \int \mathrm d m \ \bm{\xi}^*_i \cdot \left[{1 \over \mu_0}\left((\nabla \times \mathbf{B})\times \mathbf{b}_j - \mathbf{B} \times (\nabla \times \mathbf{b}_j)\right) - {1 \over \rho}{\nabla \cdot (\rho \bm{\xi}_j)}\ \mathbf{B} \times \nabla \times \mathbf{B}\right],
\end{equation}
and
\begin{equation}
    \mathbf{b}_j = \nabla \times \left(\bm{\xi}_j \times \mathbf{B}\right).
\end{equation}
This matrix generalises the linear expressions in \citet[][their eqs. 18-21]{bugnet_magnetic_2021}, which yield its diagonal elements. Accordingly, by a similar argument to the one we have prosecuted above for pure rotation, we can conclude that the calculations performed in that work are correct only to leading order in \(\kappa\). Likewise, we therefore expect any deviations from these first-order calculations to also be most significant in the vicinity of the p-dominated mixed modes, where near-degeneracy effects emerge.

\begin{figure}[htbp]
    \centering
    \annotate{\includegraphics[width=.475\textwidth, trim=.25cm .25cm .25cm .25cm,clip]{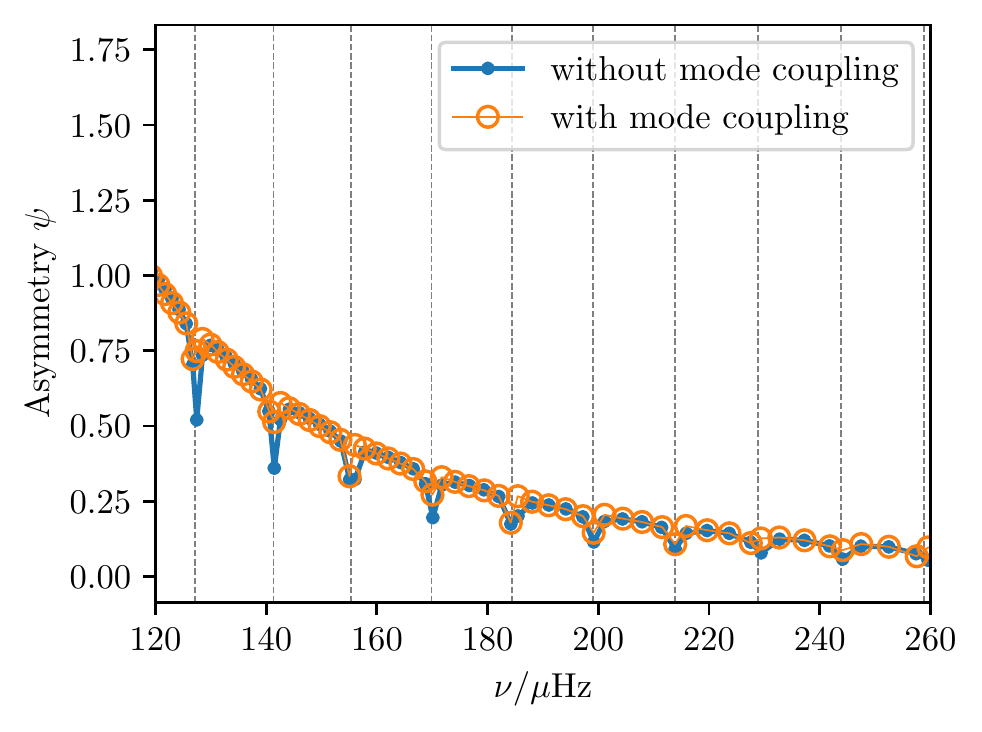}}{\node at (.87, .5){\textbf{(a)}};}
    \annotate{\includegraphics[width=.495\textwidth, trim=.25cm .25cm .25cm .25cm,clip]{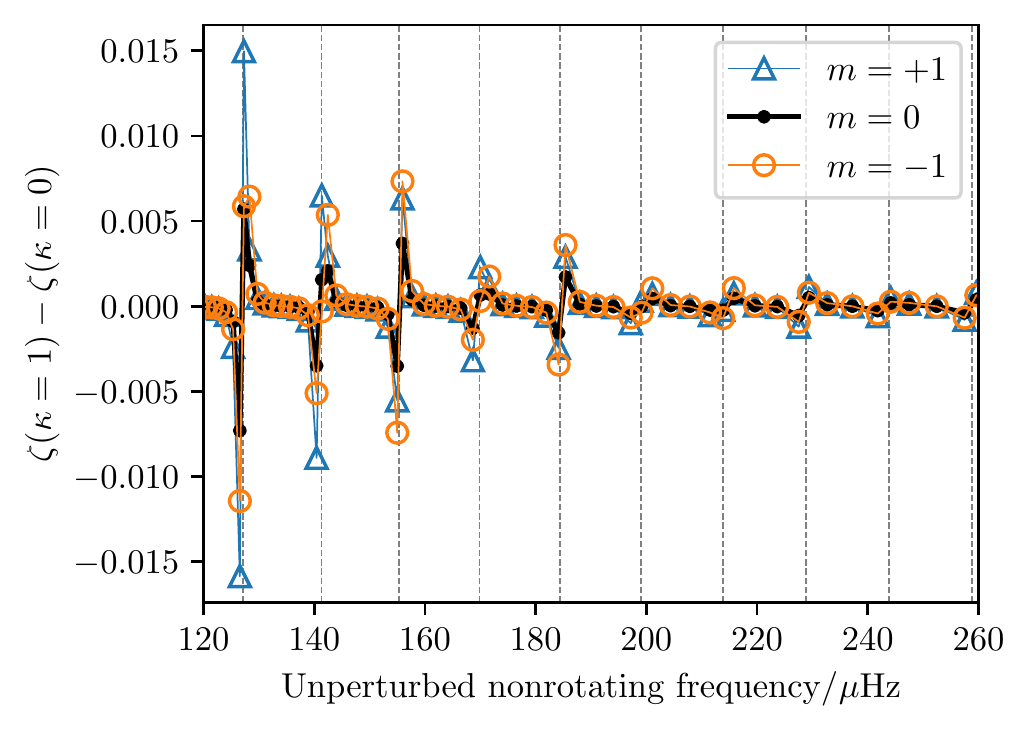}}{\node at (.9, .22){\textbf{(b)}};}
    \caption{Magnetic asymmetries induced on the mixed-mode dipolar multiplets for Model 1 when considering a field of amplitude 1MG, as done in \citet{bugnet_magnetic_2021}. Blue line and points indicate the magnetic perturbation computed on mixed-mode frequencies in presence of mode coupling, and the orange circle shows the result when mode coupling is not taken into account. The locations of the $\pi$ modes are marked out with the vertical dashed lines in both panels. \textbf{(a)} Comparison of multiplet asymmetries computed with only the first-order expressions, vs. those emerging from the full mode-coupling calculations. \textbf{(b)} Differences in $\zeta$ between modes with and without accounting for the magnetic field perturbation --- these differences are ignored in the first-order construction.}
    \label{fig:mag}
\end{figure}

We use Model 1 again to illustrate the differences between the linear and general approach to mode coupling in the presence of magnetism. 
Accordingly, we have two different sets of asymmetry parameters (evaluated with respect to the linear and general approach of mode coupling), which we compare in \cref{fig:mag}a. As in \citet{bugnet_magnetic_2021}, we see that the first-order approach of mode coupling (shown with blue points and lines) gives much larger values of the asymmetry parameter for g-dominated mixed modes of than it does for p-dominated mixed modes, with frequencies near those of the underlying \(\pi\) modes. When mode coupling is taken into account (orange lines and open circles), the multiplet asymmetries for g-dominated mixed modes are in good agreement with those returned from the first-order expression. However, those of the p-dominated mixed modes exhibit differences from the first-order expressions that are numerically much larger than the resonant asymmetries resulting from pure rotation (as in \cref{fig:asym1}a). This is especially evident at low frequencies, where the density of modes is highest. As such, when diagnosing magnetism inside red giants using the asymmetry parameter, it is important to limit the study to g-dominated modes, to avoid false-positive magnetic detections induced by mode coupling asymmetries.

In our discussion above, we demonstrated that asymmetric splitting becomes significant when the mixing fraction $\zeta$ differs significantly between components of the same multiplet. A similar phenomenon can be identified here: interactions between mode coupling and the magnetic perturbation are most significant when the magnetic perturbation induces differential changes to the mixing fractions $\zeta$ as well. We demonstrate this in \cref{fig:mag}b, where we show the changes to the mixing fraction which can be attributed to the action of the magnetic field (interacting with mixed-mode coupling). Multiplets in \cref{fig:mag}a where the linear and nonlinear asymmetries diverge most correspond to those in \cref{fig:mag}b where the change in $\zeta$ induced by the magnetic perturbation is most different between multiplet components.

\begin{figure}
    \centering
    \includegraphics{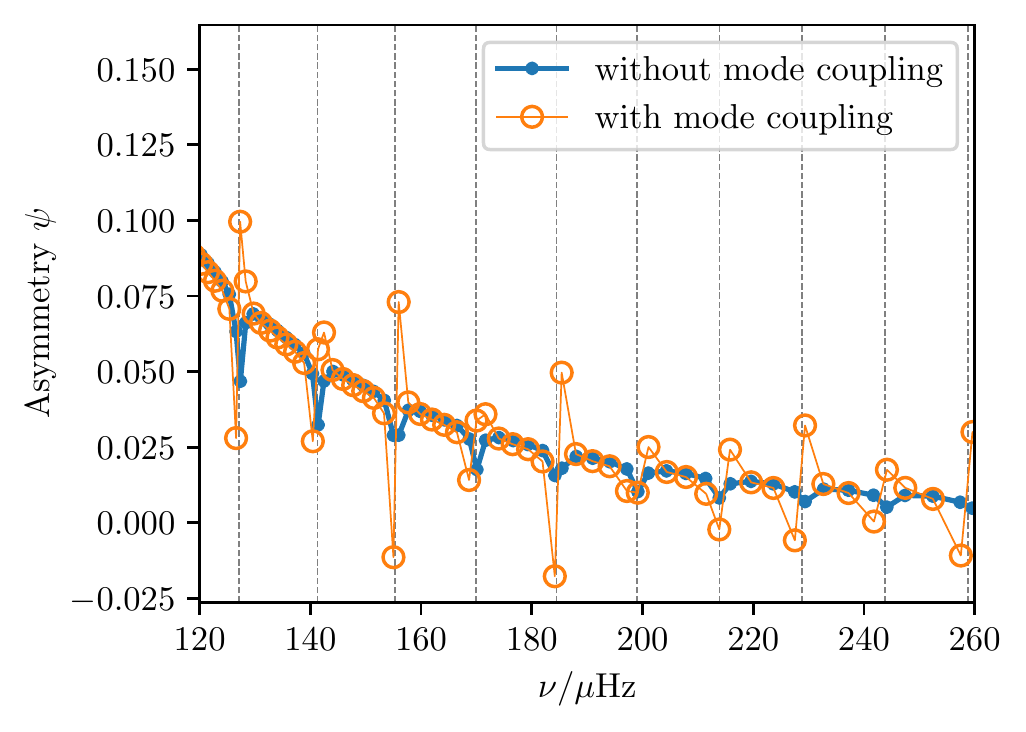}
    \caption{Same quantities as \cref{fig:mag}, but with a magnetic field strength of 0.3 MG.}
    \label{fig:mag2}
\end{figure}

In summary, the linear description of asymmetric splittings arising from magnetic fields appears to hold very well for the most g-dominated mixed modes. Importantly, we also demonstrate that rotational coupling effects do not significantly affect the characteristic magnetic signature varying in $1/\nu^3$ for g-dominated modes \citep{bugnet_magnetic_2021}. We therefore insist on the importance of the accurate selection of sufficiently g-dominated mixed modes for the search for magnetic signatures inside red giants.

However, for the near-resonance multiplets (in the p-dominated mode regions), accounting for mode coupling yields deviations from this linear description that are potentially substantially larger than the intrinsic values associated with pure rotation. These ultimately are a consequence of interactions between rotational splitting, mode coupling, and the magnetic perturbation near p-dominated modes, which are not immediately apparent when these phenomena are examined in isolation. While the first-order expression may suffice for a rough estimate of the field strength, detailed characterisation on a multiplet-by-multiplet basis will require both rotation and the magnetic field to be modelled simultaneously; this is known already to be necessary when treating other aspects of the interaction between rotation and magnetism \citep[e.g.][]{loi_topology_2021}.

Finally, we note that the morphology of this behaviour is highly dependent on the strength of the magnetic field. We show in \cref{fig:mag2} the asymmetry parameters obtained using the same magnetic configuration, but with the field strength scaled down to $0.3\ \mathrm{MG}$. The size of the associated frequency perturbation scales with $B^2$, so this results in asymmetry parameters on the g-dominated multiplets that are an order of magnitude smaller than in \cref{fig:mag}. However, the near-resonance multiplets exhibit \edit1{deviations} from this smooth curve of the kind seen in \cref{fig:asym1}a.

\hypertarget{evolutionary-considerations}{%
\subsection{Evolutionary Considerations}\label{evolutionary-considerations}}

We have now derived expressions, \cref{eq:psi:mixbound2,eq:eps:mixbound2}, to estimate both the resonant asymmetry in the rotational splitting, as well as the systematic error in the true rotational width from that given by the first-order expression, \cref{eq:zetarot}. In order to illustrate how these may change over the course of stellar evolution, we must additionally specify how \(\Omega_\text{env}\) and \(\Omega_\text{core}\) evolve over time in a reasonably realistic fashion. For this purpose, we construct evolutionary models including rotation using the angular momentum transport prescription of \citet{fuller_slowing_2019}. As in that work, we choose initial conditions of solid-body rotation on the zero-age main sequence. The central result of that work is that core rotation rates during first ascent up the red giant branch are significantly more sensitive to an angular momentum transport efficiency parameter \(\alpha\) than to the initial rotational period. Accordingly, we use \(P_{\text{rot},\text{ZAMS}} = 2\ \mathrm{d}\), as in that work, for a series of evolutionary tracks with initial mass going from \(1.4\) to \(2.0\ M_\odot\). In a concession to verisimilitude, we set \(P_{\text{rot},\text{ZAMS}} = 20\ \mathrm{d}\) for a further evolutionary track with initial mass \(1.2\ M_\odot\), and \(P_{\text{rot},\text{ZAMS}} = 30\ \mathrm{d}\) for one at \(1.0 M_\odot\), since the mechanism of \citet{fuller_slowing_2019} does not include main-sequence magnetic braking.

\begin{figure}[htbp]
    \centering
    \annotate{\includegraphics[width=.485\textwidth, trim=.25cm .25cm .25cm .25cm,clip]{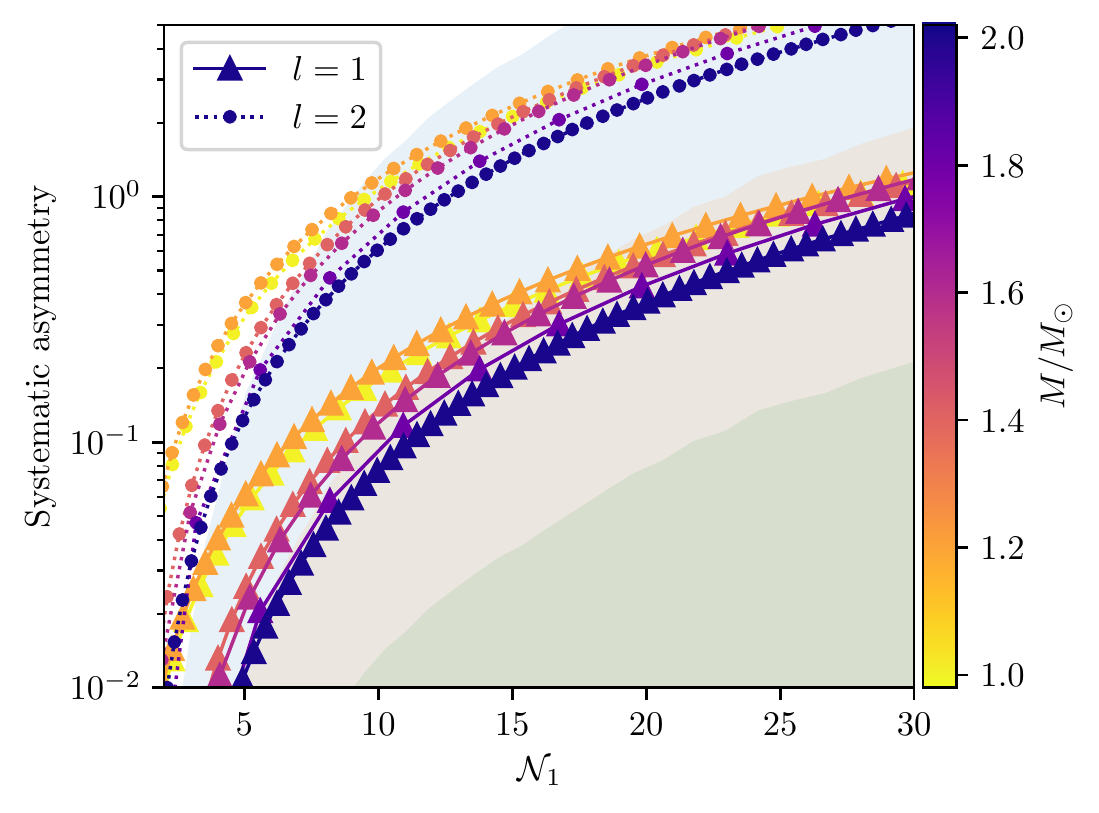}}{\node[left] at (.8, .2){\textbf{(a)}};}
    \annotate{\includegraphics[width=.485\textwidth, trim=.25cm .25cm .25cm .25cm,clip]{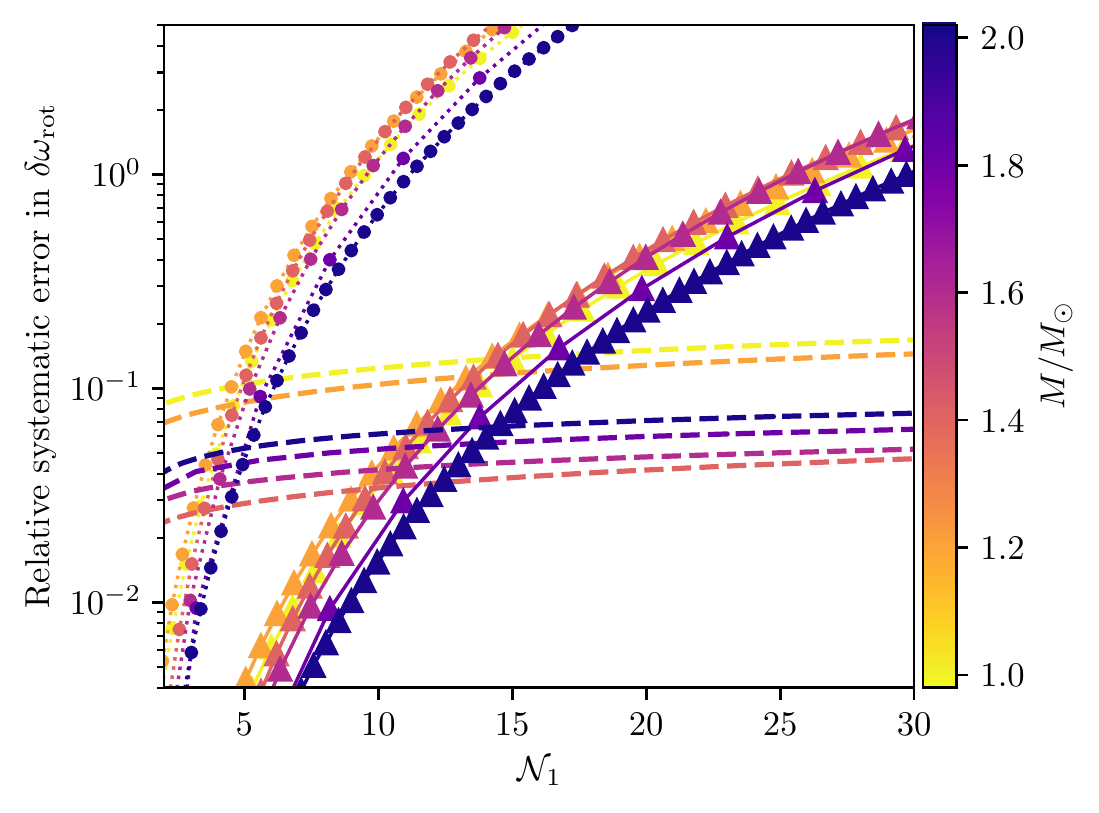}}{\node[left] at (.25, .9){\textbf{(b)}};}
    \caption{Systematic asymmetry \textbf{(a)} and relative errors \textbf{(b)} induced into pure rotational splitting from multimode coupling, over the course of post-main-sequence evolution, for MESA models with angular momentum transport (see text for description). All quantities are evaluated at \numax. Both are shown with respect to $\mathcal{N}_1$, the number of dipole g-modes per p-mode. Colours show different stellar masses, with different initial conditions (see text for complete description). The shaded regions in \textbf{(a)} show upper limits on the magnetic asymmetry in pure g-modes multiplets in presence of a fossil magnetic field. We show these limits with a field strength of 1 MG (blue zone), 0.3 MG (gray zone), and 0.1 MG (green zone) at Model 1 ($\mathcal{N} \sim 5$). \textbf{(b)} shows the relative statistical error associated with an absolute measurement uncertainty of $10\ \mathrm{nHz}$ \cite[as in][]{gehan_core_2018}.}
    \label{fig:evol}
\end{figure}

For post-main-sequence mixed-mode oscillators, we relate their rotational profiles to the two-zone model of differential rotation by evaluating averaged core and envelope rotation rates as
\begin{equation}
    \Omega_\text{core} \sim {\int_0^{r_\text{cz}} \Omega(r) N(r) / r \ \mathrm d r \over \int_0^{r_\text{cz}} N(r) / r \ \mathrm d r},\ \ \ \Omega_\text{env} \sim {\int_{r_\text{cz}}^R \Omega(r) / c_s \ \mathrm d r \over \int_{r_\text{cz}}^R 1 / c_s \ \mathrm d r},
\end{equation}
using the WKB expressions for g- and p-mode wavenumbers, respectively. We insert these into \cref{eq:psi:mixbound2,eq:eps:mixbound2}, and show the evolution of these estimators, as evaluated at \numax, in \cref{fig:evol}. For comparison with earlier works \citep{gehan_core_2018, bugnet_magnetic_2021}, we display them as functions of \(\mathcal{N}_1 = \Delta\nu / \nu^2 \Delta\Pi_1\), the number of dipole g-modes per dipole p-mode.

Specifically, \cref{fig:evol}a shows the systematic splitting asymmetry induced from near-degeneracy effects. Of particular concern is that these asymmetries have been proposed for use as diagnostic measurements of magnetic signatures in evolved stellar cores \citep[e.g.][]{bugnet_magnetic_2021}. For comparison, we also mark out with the blue, gray, and green zones the range of dipole-mode asymmetries anticipated to arise from such magnetic effects described in that work, and neglecting the effects of mode coupling, with a field strength of 1 MG, 0.3 MG, and 0.1 MG at Model 1, respectively. These regions are bounded from above by the linear asymmetries from g-modes (evaluated using the dipole $\gamma$-mode kernels nearest to $\numax$). The field strength is scaled with the size of the radiative core to conserve the magnetic flux inside the radiative region along the evolution, but no further dissipating mechanisms are considered. Roughly speaking, we see that both sources of asymmetric splitting evolve in roughly the same fashion with increasing $\mathcal{N}$. As such, isolation of the two effects can only be effected by selecting either g- or p-dominated multiplets (for magnetic signatures vs. rotational coupling), rather than e.g. preferentially selecting more- or less-evolved targets for observation.

Moreover, the magnetic asymmetries are also known to decrease with increasing azimuthal degree. Thus, we see also that the size of the quadrupole-mode intrinsic rotational asymmetry remains larger than those resulting from magnetic fields, within the range of evolution and field strengths that we have considered. This serves as a further reason \citep[in addition to those provided in][]{bugnet_magnetic_2021} to prefer diagnoses of magnetic fields as made with dipole modes over those made using quadrupole modes.

In summary, were they not accounted for, these degeneracy-induced asymmetries in near-resonance multiplets may well lead to spurious diagnoses of magnetic signatures indicating magnetic fields much stronger than would actually exist. This risk would be exacerbated if such measurements were to be made from quadrupole modes, and is most easily avoided with observational access to \edit1{the most} g-dominated multiplets.

\cref{fig:evol}b shows the evolution of \cref{eq:eps:mixbound2}, the systematic error from the first-order expressions. Generally speaking, \cref{eq:psi:mixbound2,eq:eps:mixbound2} are proportional to powers of \(\mathcal{N}_\text{rot} = m\Omega / 2\pi \nu^2 \Delta\Pi\), which (roughly speaking) may be interpreted as the number of g-modes per rotational splitting width. As we have seen, when \(\mathcal{N}_\text{rot}\) is larger than 1, crude nearest-neighbour identification of rotational multiplets fails, and measurements of rotational widths require some other, more sophisticated approach to mode identification. However, these methods still rely on the first-order expressions \citep[e.g.][]{mosser_spin_2012, gehan_core_2018}, and therefore remain susceptible to systematic errors from inadequate treatments of avoided crossings. Measurements made using these techniques in these existing works lie within the range \(\mathcal{N}_1 \lesssim 30\). We mark out with dashed lines in \cref{fig:evol}b the nominal statistical errors of these methods corresponding to their reported absolute measurement errors in the frequency widths. We nonetheless find that the systematic error induced into these methods as a result of ignoring higher-order mode-coupling effects is potentially much larger than their reported errors, within the range of \(\mathcal{N}_1\) in which they have been applied.

\hypertarget{prospects-for-the-inverse-problem}{%
\section{Prospects for the Inverse Problem}\label{prospects-for-the-inverse-problem}}

\label{sec:prospects}

We have shown above that even in the two-zone model, the use of the first-order expression, \cref{eq:zetarot}, may systematically misestimate the true sizes of the splittings associated with mixed-mode rotational multiplets, given a particular configuration of core and envelope rotation rates. We stress that this does not alter our ability to actually measure these multiplet widths (e.g.~Ong \& Gehan in prep.); rather, these systematic errors may interfere with our interpretation of the measured quantities as being averaged rotation rates specified by \cref{eq:kern}. We therefore seek generalisations of this expression to use for solving the rotational inverse problem, both in the two-zone model, as well as for more advanced rotational inversion techniques.

\hypertarget{splittings-in-the-pigamma-basis}{%
\subsection{\texorpdfstring{Splittings in the \(\pi\)/\(\gamma\) basis}{Splittings in the \textbackslash pi/\textbackslash gamma basis}}\label{splittings-in-the-pigamma-basis}}

The isolated basis sets of \(\pi\) and \(\gamma\) modes are an alternative formulation of describing mixed modes, in lieu of the natural mixed-mode eigenfunctions. While the wave operator is itself not diagonal in this basis, its off-diagonal coupling elements can be easily computed with respect to a stellar model (e.g.~using the construction of \ob). \edit1{In \autoref{subsec:perturbation}, we made the observation that modifying perturbation theory to explicitly account for near-degeneracy effects would require us to diagonalise the restriction of the rotation operator to within each near-degenerate subspace. We have now shown from both analytic considerations and numerical results that the rotation matrices in a star exhibiting mixed modes have significant off-diagonal structure in the natural basis of mixed modes, but are very well-approximated as being diagonal in the isolated basis of \(\pi\) and \(\gamma\) modes. Correspondingly, we may interpret the $\pi$ and $\gamma$ modes to be the natural choice of basis functions through which rotation lifts the degeneracy on each of these subspaces.}

\edit2{Moreover, we have shown that for near-degenerate multiplets, the linear response of mixed modes to rotation, \cref{eq:zetarot}, must (generally speaking) be extended to include contributions from off-diagonal matrix elements of the rotation operator. By contrast, since the rotation matrix is close to diagonal in the isolated basis of p/g modes, this linear treatment remains good there even near degeneracy; any nonlinear behaviour in the mixed modes arises from the coupling between the two mode cavities, which does not depend on the rotational configuration of the star.} This being the case, \edit1{we suggest the use of a basis of isolated p/g-modes for characterising stellar rotation in the two-zone model.}

\edit1{While the splittings of these notional pure modes cannot be directly observed,} we propose a procedure by which the mean core (and potentially envelope) rotation rates may still be constrained in a least-squares sense. Given a set of \edit1{pure} p- and g-mode frequencies and the coupling between them, we may define rotationally-split mixed-mode frequencies as the eigenvalues of the QHEP:
\begin{equation}
    \left(
    \omega^2
    \begin{bmatrix}
    \mathbb{I}_p & \mathbf{D} \\
    \mathbf{D}^T & \mathbb{I}_g
    \end{bmatrix}
    +
    2 m \omega
    \begin{bmatrix}
    \beta_\pi \Omega_\text{env}\mathbb{I}_p & 0 \\
    0 & \beta_\gamma \Omega_\text{core}\mathbb{I}_g
    \end{bmatrix}
    +
    \begin{bmatrix}
    \mathbf{\Omega}_p^2 & \mathbf{A} \\
    \mathbf{A}^T & \mathbf{\Omega}_g^2
    \end{bmatrix}
    \right)\mathbf{c} = 0, \label{eq:matrixls}
\end{equation}
in block matrix form, where \(\mathbf{\Omega}_g\) and \(\mathbf{\Omega}_p\) are diagonal matrices containing the nonrotating g- and p-mode angular frequencies, and \(\mathbf{A}\) and \(\mathbf{D}\) are the coupling coefficients between the \(\pi\) and \(\gamma\)-mode basis functions. In particular, operating within the \(\pi/\gamma\) construction permits us to neglect the off-diagonal entries of the second matrix in this problem. Supposing that the other quantities entering into the problem may be adequately constrained by stellar modelling and/or the \(m=0\) mixed-mode frequencies, the rotating mixed-mode frequencies will then only depend on two additional parameters, which are \(\Omega_\text{env}\) and \(\Omega_\text{core}\). This gives a generative model for the rotating mode frequencies, such that the parameters \(\Omega_\text{env}\) and \(\Omega_\text{core}\) may be constrained from the data in the usual fashion (e.g.~by \(\chi^2\)-minimisation/likelihood maximisation).

\hypertarget{beyond-two-zone-differential-rotation}{%
\subsection{Beyond Two-Zone Differential Rotation}\label{beyond-two-zone-differential-rotation}}

More generally, existing analysis beyond the two-zone model of radial differential rotation, via rotational inversion techniques \citep[as employed in e.g.][]{eggenberger_asteroseismology_2019, ahlborn_rotation_2020, fellay_asteroseismology_2021}, rely on discretisations of integral kernel expressions of the form of \cref{eq:kern} --- i.e.~solutions to Fredholm equations of the first kind \citep{hansen_numerical_1992} --- with no cross terms between modes; they are therefore valid only in the same regime as the first-order expression \cref{eq:zetarot}, where the rotational asymmetry arising from avoided crossings may be neglected. Accordingly, their direct application \edit1{to where these avoided crossings cannot be neglected}, particularly to red giants, may not be correct.

\edit1{Were rotational inversions to be performed in the mixed-mode basis, our discussion of \cref{fig:off_diagonal,fig:dominance} moreover implies that we must also include the off-diagonal rotation matrix elements in the inversion problem in addition to the rotational widths, or else we would lose information about the rotational configuration in the presence of multiplet asymmetry --- they, too, depend on the rotational profile. However, we note that almost no previous attempts at rotational characterisation have explicitly measured these off-diagonal matrix elements, let alone used them in the inversion procedure. Aside from \dheuv, the only observational efforts to account for asymmetric splittings have been based on the asymptotic construction, which we have shown in \autoref{sec:construction} not to yield correct values for these off-diagonal elements even in the two-zone model of differential rotation, in any case.} \edit2{These difficulties are further compounded by the fact that, in the regime where second- and higher-order effects in the rotational splitting become significant, the dependence of the sensitivity kernels themselves on the rotation rate can also no longer be ignored, rendering the linear inversion construction itself potentially questionable.}

\edit1{Again, we propose that these inversions be carried out in the isolated basis of $\pi$ and $\gamma$ modes instead. \edit2{As we have discussed previously, a linear treatment of rotation in the basis of these isolated modes remains valid even where it does not for the associated near-degenerate mixed modes}. That the off-diagonal matrix elements $\alpha_{ij}$ and $D_{ij}$ must be specified is not a significant methodological complication, as they do not depend on the rotational properties of the star. Instead, they can be found independently of rotation, through matching the observed $m=0$ modes to constrain the structure of the star. The stellar structure must be constrained well in this fashion in order to produce a good enough fiducial model for rotational inversions to be feasible in the first place, so these parameters will already be available from the fiducial model, and do not enter as unknowns into the rotational inverse problem. Whereas working in the mixed-mode basis would require both the diagonal and off-diagonal elements of the rotation matrix to be specified from a fixed set of modes, by working in the $\pi/\gamma$-mode basis, and approximating the rotation matrix as being diagonal, we essentially impose a sparsity constraint on the inferred rotation matrix for free (having already specified the fiducial structure); the extra information goes into reducing the statistical uncertainty on the inferred quantities. Finally, the surface term must be corrected for in deriving the fiducial structure for inversion kernels; new surface term corrections for mixed modes operate in the basis of $\pi$ and $\gamma$ modes in any case \citep[e.g.][]{ong_surface_1,ong_surface_2}, thus simplifying matters were it to also be used for the inversion procedure.}

Our discussion above illuminates how a linear inversion procedure may be recovered in \edit1{the basis of $\pi$ and $\gamma$ modes}. In particular, we may replace the second term in \cref{eq:matrixls} with a diagonal matrix as \(2 m \omega \mathbf{R}\), and constrain its entries in e.g.~the usual least-squares fashion. We note that this procedure remains well-posed: for \(N\) dipole-mode multiplets under consideration, there are \(N\) diagonal entries in this matrix \(\mathbf{R}\), but \(2N\) rotationally-split sectoral multiplet components against which they are to be constrained. These diagonal matrix elements then specify the independent variables \(\delta\omega_\text{rot}\) of integral equations of the form of \cref{eq:kern}, but with the relevant integral kernels being associated with the underlying isolated basis of \(\pi\) and \(\gamma\) modes, rather than the mixed modes directly.

\begin{figure*}[htbp]
    \centering
    \annotate{\includegraphics[width=.485\textwidth, trim=.25cm .25cm .25cm .25cm,clip]{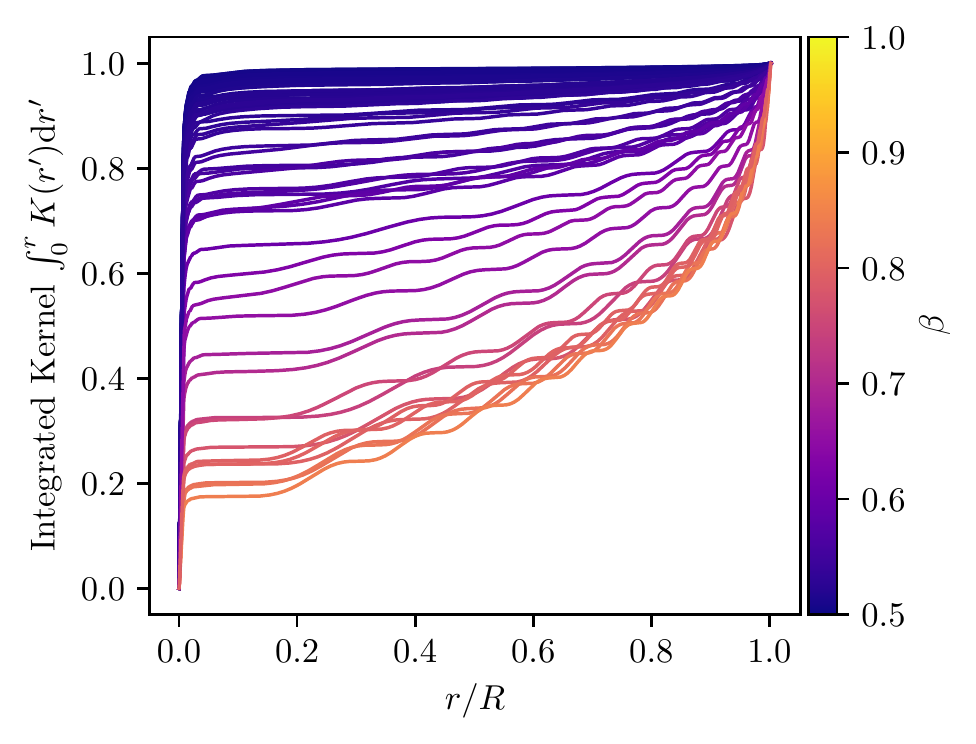}}{\node at (.77, .23){\textbf{(a)}};}
    \annotate{\includegraphics[width=.485\textwidth, trim=.25cm .25cm .25cm .25cm,clip]{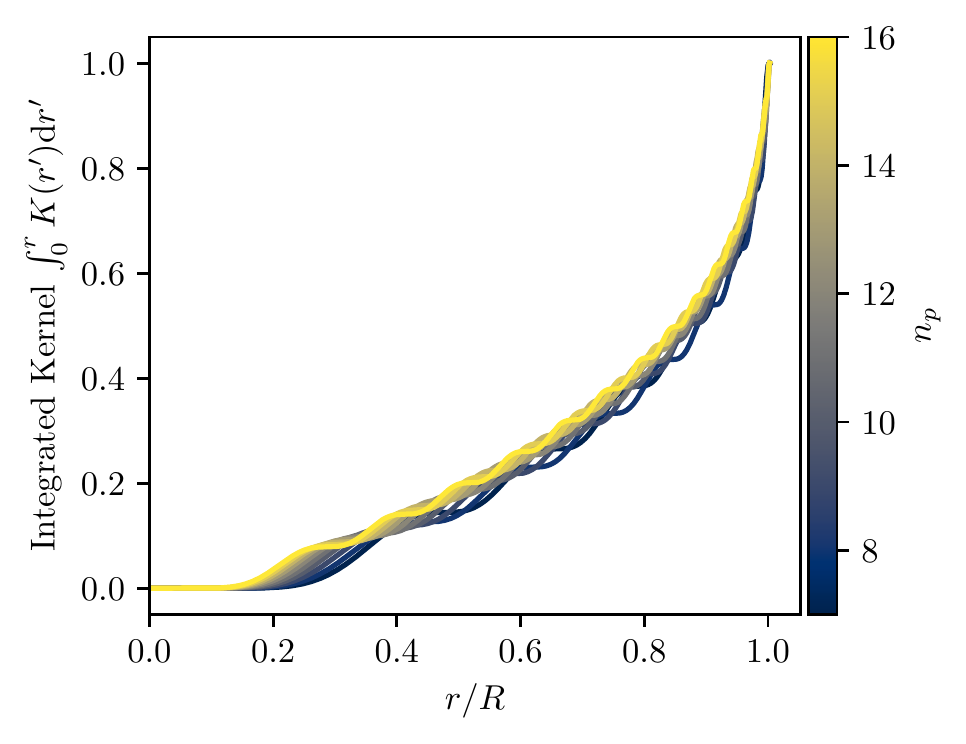}}{\node at (.77, .23){\textbf{(b)}};}
    \annotate{\includegraphics[width=.485\textwidth, trim=.25cm .25cm .25cm .25cm,clip]{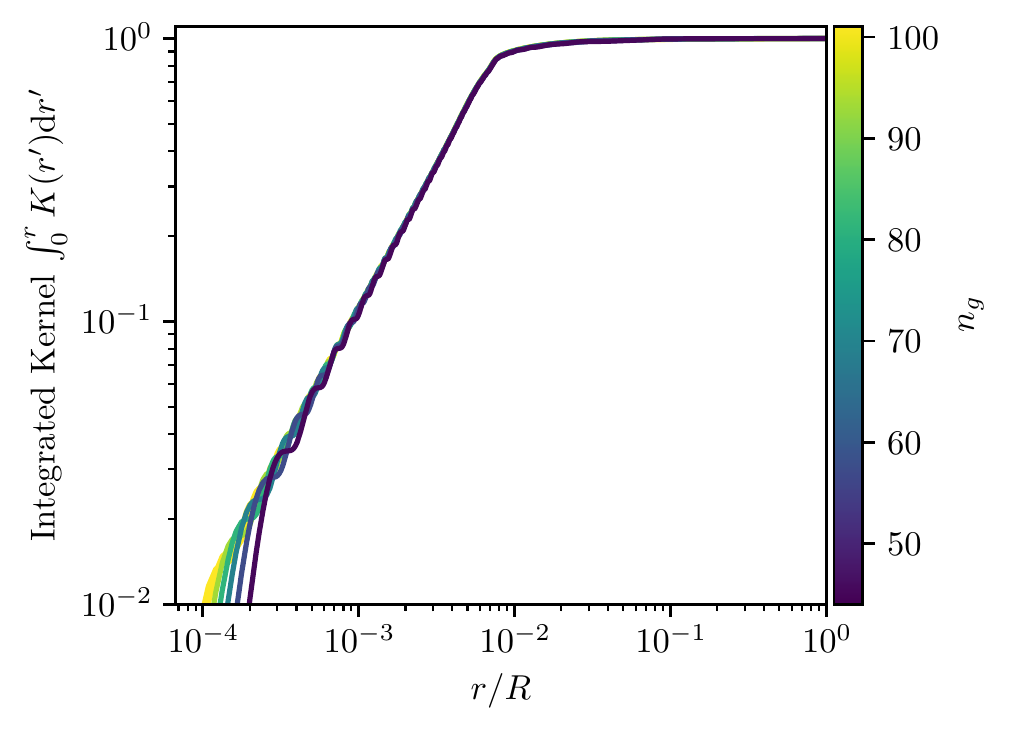}}{\node at (.77, .23){\textbf{(c)}};}
    \caption{Cumulative integrals of different families of rotational inversion kernels associated with dipole modes, all in the same evolved stellar model. \textbf{(a)} Standard inversion kernels computed using \cref{eq:rotkernel} with respect to mixed-mode eigenfunctions, as typically used in the literature. These exhibit sensitivity to both the radiative core and to the convective exterior, to differing extents depending on the mixing fraction $\zeta$ of the modes under consideration.  \textbf{(b)} Rotational kernels from $\pi$-mode eigenfunctions. Since dipole $\pi$-modes have an interior turning point at $\numax$ exterior to the radiative core, these inversion kernels are essentially insensitive to the core, unlike the mixed-mode kernels. \textbf{(c)} Rotational kernels from $\gamma$-mode eigenfunctions. The localisation of these kernel functions is such that they decay exponentially outside of the compact radiative core.}
    \label{fig:kernels}
\end{figure*}

We illustrate the differences between these kinds of integral kernels in \cref{fig:kernels}, in which we plot the cumulative integral \(I(r) = \int_0^r K(r') \mathrm d r'\) associated with each integral kernel \(K\), as computed with respect to Model 1 using \cref{eq:rotkernel}; the results for Model 2 are qualitatively very similar. In \cref{fig:kernels}a, we show these cumulative integrals with respect to the usual basis set of normal modes, as is typically done. As a consequence of mode mixing, all of these modes are to some extent sensitive to differential rotation in both the compact radiative core and the diffuse convective envelope. The most g-dominated mixed modes, with \(\beta \sim 1/2 \iff \zeta \sim 1\), are minimally sensitive to the envelope, while in principle a strictly p-dominated mixed mode with \(\beta \sim 1 \iff \zeta \sim 0\) would be minimally sensitive to the core. In these red giants, however, the configuration of the mode cavities is such that there is always at least one dipole g-mode close to resonance with every dipole p-mode, and so no such nearly-pure p-dominated mixed modes can exist.

However, the procedure we have elucidated above permits us to operate with respect to notional rotational splittings associated separately with \(\pi\) and \(\gamma\)-mode basis functions, which have their own rotational kernels in lieu of those computed from the mixed modes. In \cref{fig:kernels}b, we show the cumulative integrals of kernels derived from the \(\pi\)-mode eigenfunctions. Unlike those of the mixed modes, these kernel functions are entirely insensitive to the core, as the \(\pi\) modes do not propagate in the interior radiative cavity if it lies entirely within the inner turning point set by the dipolar Lamb frequency. Likewise, in \cref{fig:kernels}c we show the same quantities for rotational kernels derived from \(\gamma\)-mode eigenfunctions. Here we see that these integrals are flat (and therefore the localisation kernels vanish) outside of the boundary of the radiative core at \(\sim r = 10^{-2} R\).

As such, if rotational splittings can be individually assigned to \(\pi\) and \(\gamma\) modes, e.g.~using the least-squares procedure that we have described above, then rotational inversion may be performed with respect to sets of basis kernel functions that have some desirable localisation properties: constraints on differential rotation may be localised to either entirely within the convective envelope, or the radiative core. This stands in contrast to the use of mixed-mode kernels, as are currently used for rotational inversions: the use of these isolated kernels may assist in allieviating shortcomings of existing techniques.

For example, \edit1{in the case of envelope rotation}, the dependence of essentially all mixed-mode kernels on the core rotation rate renders the problem of localising rotational measurements to the envelope \edit1{to be very poorly numerically conditioned \cite[e.g.][]{deheuvels_seismic_2012,deheuvels_seismic_2014,ahlborn_rotation_2020}. Even the most recent advances in technique \edit2{\citep{ahlborn_improved_2022}} produce at most a single, maximally core-insensitive rotation kernel from a given set of modes, precluding the localisation of differential rotation in particular. These numerical difficulties} may be circumvented by the use of \(\pi\)-mode kernels, which are already insensitive to the core. \edit1{Moreover, our numerical construction recovers exactly as many $\pi$-mode kernels as there are p-dominated mixed modes, thereby maximising the use of the available information. In the case of core rotation,} \citet{fellay_asteroseismology_2021} assert to have constrained the shape of radial differential rotation localised to \edit1{near} the boundary of the radiative core --- in tension with the claim made in \citet{wilson_rotational_2021} that no meaningful constraints of this kind are possible. The use of \(\gamma\)-mode kernels here may refine these arguments and assist in resolving this tension, by eliminating any potential interference owing to envelope rotation. 

\edit2{We n}ote that \edit1{some of} these studies \edit1{of both core and envelope rotation} have been performed on subgiants, with low \(\mathcal{N}_1(\numax)\), where nominally the first-order expressions remain valid, per \autoref{sec:numerics}. While this means that the use of isolated \(\pi/\gamma\) kernels here is not mandatory, our discussion here suggests that it might nonetheless be advantageous compared to conventional methods. \edit2{Moreover, we must also note that these attempts at localising rotational measurements have so far been effectively restricted to the two-zone model of radial differential rotation (i.e. attempting to decouple the core from the envelope), as generalisations to this have thus far not been considered feasible owing to the abovementioned methodological difficulties \cite[in particular their appendix D]{ahlborn_improved_2022}. Having decoupled the two mode cavities, we are now at least in principle equipped to explore prospects for estimating rotation rates and gradients at specific target locations, as ordinarily done in helioseismology \citep[e.g.][]{pijpers_sola_1994}.  We defer a more detailed examination of the localisation properties of $\pi$/$\gamma$ mode kernels to a later work.}

\hypertarget{conclusion}{%
\section{Conclusion}\label{conclusion}}

Existing techniques for internal rotational characterisation of evolved stars require the rotation operator to be effectively diagonal with respect to the standard functional basis of normal modes --- or equivalently, in the two-zone model of radial differential rotation, that the observed multiplet splitting for any given mixed mode be a direct linear combination of contributions from the core and envelope. However, this linearity assumption is known not to hold for mixed modes near resonance in evolved subgiant stars, as a result of avoided crossings between the underlying p- and g-modes occurring at different ages for different multiplet components \citep{deheuvels_near_2017}. While this result has raised questions regarding the correctness of existing rotational characterisations made using this linearity assumption, the manner in which said assumption might break down over the course of stellar evolution has not previously been investigated in detail.

A fundamental limitation of earlier investigation of this phenomenon was the restricted \edit1{development} of the analytic theory to where avoided crossings in near-degenerate mixed modes could be treated in isolation. Previous studies have been limited to the case of p- and g-modes interacting in isolated pairs in subgiants, rather than long-range many-to-one mode coupling of the kind potentially seen in red giants. In this work, we \edit1{operate} with respect to a different set of basis functions than the usual normal modes of oscillation --- in particular, we use the \(\pi/\gamma\) decomposition of \citet{ong_semianalytic_2020}. By doing so, we are able to not only demonstrate that a breakdown of linearity does indeed occur in the regime of dense-g-sparse-p mode mixing in evolved red giants, but also examine the detailed dynamics of how this occurs. Our linear-algebraic treatment yields rich rotationally-modulated families of avoided crossings (e.g.~\cref{fig:avoidedcrossings}) which are qualitatively consistent with those seen to emerge from nonperturbative rotating pulsation calculations \citep{ouazzani_rotation_2013}. \edit2{A linear treatment of rotation remains applicable in the decoupled basis of isolated p- and g-modes even when it does not for the mixed modes.}. The use of these decoupled basis functions also illuminates intrinsic contradictions in the most commonly used alternative approach --- that of the asymptotic mode-bumping function \(\zeta\) --- either when relaxing the requirement that \(\Omega_\text{env} = 0\), or more generally when leaving the two-zone model of radial differential rotation.

From perturbation analysis of the associated matrix eigenvalue problem, we derive analytic estimators for both the asymmetry and systematic error in the rotational splitting owing to near-degeneracy effects, which, in conjunction with stellar modelling under some coarse assumptions about angular momentum transport in evolved stars, permit us to roughly constrain the regime of evolution in which the linear expressions hold well. In particular, we find that the direct interpretation of existing measurements of dipole-mode rotational splitting as core rotation rates becomes increasingly questionable for \(\mathcal{N}_1(\numax) \gtrsim 10\), and that in such cases the rotational asymmetry induced by mode coupling also becomes significant. This is well within the limits of evolutionary states in the present observational sample of rotating red giants for which measurements of rotation rates have been made. In this r\'egime, these near-degeneracy effects can also be shown to dominate the intrinsic asymmetry and systematic error arising from higher-order effects of rotation, such as centrifugal forces and structural deformation. We have moreover briefly examined how these near-degeneracy effects interact with the presence of internal magnetic fields which produce frequency shifts that also depend on the azimuthal order \(m\). We conclude that while magnetic field effects largely dominate the asymmetry of g-dominated multiplets, magnetic and mode coupling asymmetries might have similar amplitude near the nominal p mode.

Finally, we have described prescriptions for how one may infer purely symmetric rotational splittings given potentially asymmetric observed multiplets, using the fact that the rotation matrix remains approximately diagonal in the isolated basis of \(\pi\) and \(\gamma\) modes for such stars. As a further benefit, existing descriptions of radial differential rotation beyond the two-zone core-envelope model may also continue to be applied in this basis without loss of correctness. We further demonstrate that the use of isolated \(\pi/\gamma\) rotational kernels may be desirable even where the linearity assumption holds, owing to differences in their localisation properties compared to rotational kernels produced directly from mixed-mode eigenfunctions.

These near-resonance effects have traditionally been avoided altogether by attempting to restrict analysis to the most g-dominated mixed-mode multiplets. However, this may not always be feasible. The most g-dominated dipole mixed modes occur at frequencies comparable to those of the intervening radial p-modes and quadrupole $\pi$-modes; in any case, the amplitudes of mixed modes scale with how much p-like character they possess. This makes dealing with near-resonance phenomena unavoidable to some extent, as pure g-modes would be essentially unobservable otherwise. Ongoing efforts to manage these effects have relied on an asymptotic construction for the mode-mixing function $\zeta$, as we have described above. In the next paper in this series (Ong \& Gehan, in prep.), we will examine the robustness of this asymptotic construction, and reconcile it with the algebraic prescription that we have used in this work.

\label{sec:conclusion}


We thank B. Mosser, J. Tayar, J. van Saders, and M. Pinnsoneault for constructive comments on early versions of this manuscript. \edit2{We also thank the anonymous referee, whose highly constructive comments and suggestions have substantially improved the quality of this work. JO acknowledges support from NASA through the NASA Hubble Fellowship grant HST-HF2-51517.001 awarded by the Space Telescope Science Institute, which is operated by the Association of Universities for Research in Astronomy, Incorporated, under NASA contract NAS5-26555.}.

\software{NumPy \citep{numpy}, SciPy stack \citep{scipy}, AstroPy \citep{astropy:2013,astropy:2018}, Pandas \citep{pandas}, \mesa\ \citep{mesa_paper_1,mesa_paper_2,mesa_paper_3,mesa_paper_4,mesa_paper_5}, \gyre\ \citep{townsend_gyre_2013}.}

  \bibliography{biblio.bib}

\end{document}

%% file: macros.tex
\usepackage[capitalize]{cleveref}
\usepackage{CJKutf8}
\usepackage{newunicodechar}

\newcommand{\Dnu}{\ensuremath{\Delta\nu}}
\newcommand{\numax}{\ensuremath{{\nu_\text{max}}}}

\defcitealias{ong_semianalytic_2020}{OB20}
\defcitealias{deheuvels_near_2017}{Dh17}
\newcommand\ob{{\citetalias{ong_semianalytic_2020}}}
\newcommand\dheuv{{\citetalias{deheuvels_near_2017}}}

\newcommand\mesa{{\textsc{mesa}}}
\newcommand\gyre{{\textsc{gyre}}}

\newcommand{\chinesename}{{\begin{CJK}{UTF8}{gbsn}(王加冕)\end{CJK}}}

\DeclareRobustCommand{\okina}{%
  \raisebox{\dimexpr\fontcharht\font`A-\height}{%
    \scalebox{0.8}{`}%
  }%
}
\newunicodechar{ʻ}{\okina}

\newcommand{\annotate}[2]{\begin{tikzpicture}
    \node[anchor=south west,inner sep=0,align=center] (image) at (0,0) {
    #1
    };
    \begin{scope}[x={(image.south east)},y={(image.north west)}]
    #2
    \end{scope}
\end{tikzpicture}}
\def\bibitem{\vskip\bibskip\footnotesize\savebibitem}

\graphicspath{{./},{./figures/}}

\renewcommand{\edit}[2]{{\ifnum#1<100000%
#2%
\else%
\textbf{#2}%
\fi}}

%% file: preamble.tex

\correspondingauthor{Joel Ong}
\email{joelong@hawaii.edu}
\author[0000-0001-7664-648X]{J. M. Joel Ong \chinesename}
\affiliation{Department of Astronomy, Yale University, 52 Hillhouse Ave., New Haven, CT 06511, USA}
\affiliation{Institute for Astronomy, University of Hawaiʻi, 2680 Woodlawn Drive, Honolulu, HI 96822, USA}
\affiliation{Hubble Fellow}

\author[0000-0003-0142-4000]{Lisa Bugnet}
\affiliation{Centre for Computational Astrophysics, Flatiron Institute, Simons Foundation, 162 Fifth Ave, New York, NY 10010, USA}
\author[0000-0002-6163-3472]{Sarbani Basu}
\affiliation{Department of Astronomy, Yale University, 52 Hillhouse Ave., New Haven, CT 06511, USA}
\received{June 20, 2022}
\revised{September 20, 2022}
\accepted{October 4, 2022}
\submitjournal{\apj}
\shortauthors{Ong, Bugnet \& Basu}
\def\sectionautorefname{Section}
\def\subsectionautorefname{Section}
\def\subsubsectionautorefname{Section}

